\documentstyle[12pt]{article}
\begin{document}

\def\d{{\rm d}}
\def\p{\partial}
\def\o{\over}
\def\ie{{\it i.e.}}
\def\eg{{\it e.g.}}
\def\be{\begin{equation}}
\def\ee{\end{equation}}
\def\bea{\begin{eqnarray}}
\def\eea{\end{eqnarray}}
\def\beaa{\begin{eqnarray*}}
\def\eeaa{\end{eqnarray*}}

\def\np{Nucl. Phys.}
\def\pl{Phys. Lett.}
\def\prl{Phys. Rev. Lett.}
\def\pr{Phys. Rev.}
\def\ap{Ann. Phys.}
\def\cmp{Comm. Math. Phys.}
\def\ijmp{Int. J. Mod. Phys.}
\def\mpl{Mod. Phys. Lett.}
\def\lmp{Lett. Math. Phys.}
\def\bams{Bull. AMS}
\def\am{Ann. of Math.}
\def\jpsc{J.
 Phys. Soc. Jap.}
\def\topo{Topology}
\def\kjm{Kobe J. Math.}
\def\phyrep{Phys. Rep.}
\def\am{Adv. in Math.}

\begin{flushright}
BRX-TH-398, US-FT-39-96, BROWN-HET-1057\\
hep-th/9609116\\
September, 1996
\end{flushright}

\begin{center}
{ \Large {\bf A New Derivation of the Picard-Fuchs Equations \\
for Effective
 $N=2$ Super Yang-Mills Theories}}

\vspace{.5in}

J.M. Isidro,$^a$\footnote{Supported by Ministerio de Educaci\'on y
 Ciencia, Spain. Home address: Dep.de F\'{\i}sica de Part\'{\i}culas, 
Universidad de Santiago, 15706 Santiago, Spain.}
A. Mukherjee,$^a$\footnote{Research supported  by
the DOE under grant DE-FG02-92ER40706.}
J.P. Nunes,$^{a,b}$\footnote{Supported by Praxis XXI, Portugal and 
the DOE under
grant DE-FG0291ER40688-Task A. Current address:{\it b}.} and H.J. 
Schnitzer$^a
$\footnote{Research supported in part by
the DOE under grant DE-FG02-92ER40706.}
\end{center}

\begin{center}{\it Department of Physics, Brandeis University, 
Waltham MA
 02254-9100, USA$^a$}\footnote{{\tt isidro,
mukherjee, schnitzer@binah.cc.brandeis.edu, nunes@het.brown.edu}}\\

{\it Department  of Physics, Brown University, Providence RI
 02912, USA$^b$}
\end{center}

\abstract{ A new method to obtain the Picard--Fuchs equations of
 effective,  $N=2$ supersymmetric
gauge theories in 4 dimensions is developed. It includes both  pure
 super Yang--Mills  and
supersymmetric gauge theories with massless matter hypermultiplets. 
It applies to all classical
gauge groups, and directly produces a decoupled set of second-order,
 partial differential equations
satisfied by the period integrals of the Seiberg--Witten differential 
along the 1-cycles of the
algebraic curves describing the vacuum structure of the corresponding
 $N=2$ theory. }

\newpage

\renewcommand{\theequation}{1.\arabic{equation}}
\setcounter{equation}{0}

\section{Introduction}

There have been enormous advances in understanding the low-energy 
properties of supersymmetric
gauge theories in the last couple of years \cite{SW}. In particular, for
 $N=2$  gauge theories
with $N_f$ matter hypermultiplets,  the exact solution for the 
low-energy properties of
the Coulomb phase of the theory is given in principle by a 
hyperelliptic curve
\cite{LERCHE}--\cite{MINAHAN}. In practice, however, a great
deal of additional work is required to extract the physics embodied 
in the curve that describes the
theory in question. A given theory is characterised by a number of
 moduli \cite{SW}--\cite{MINAHAN} which
are related to the vacuum expectation values of the scalar fields of
 the
$N=2$ vector multiplet and the bare masses of the matter
 hypermultiplets. If the scalar fields of the
matter hypermultiplets have vanishing expectation values, 
then one is in the Coulomb phase of the
theory; otherwise one is in a Higgs or in a mixed phase 
\cite{SW,SEIBERG}. This paper will only be
concerned with the Coulomb phase of asymptotically free $N=2$
supersymmetric theories.

The Seiberg-Witten  (SW) period integrals 
\be
 {\vec \Pi}=\pmatrix{{\vec a}_D\cr {\vec a}\cr} 
\label{eq:zia}
\ee
are related to the prepotential \cite{Matone}
 ${\cal F}({\vec a})$ characterising 
the low-energy effective
Lagrangian by
\be
a_D^i={\p {\cal F}\o \p a_i}
\label{eq:zib} 
\ee
One can use the global monodromy properties of ${\vec \Pi}$ to 
essentially fix it, and then find
the prepotential  ${\cal F}({\vec a})$ by integration. In practice,
 one needs to construct 
the SW
periods ${\vec \Pi}$ from the known, auxiliary hyperelliptic curve;
 not an easy task for groups
with rank 2 or greater. One particular way of obtaining the 
necessary  information is to derive a
set of Picard-Fuchs (PF) equations for the SW period integrals.
The PF equations have been
formulated for $SU(2)$ and $SU(3)$ with $N_f=0$ 
\cite{KLEMM,BILAL}, and
for 
$N_f\neq 0$ for massless 
hypermultiplets $(m=0)$ \cite{ITO,RYANG}. The solutions to these
equations have been considered  for $
SU(2)$
with $N_f=0, 1, 2, 3$, for $SU(3)$ with $N_f=0, 1, \ldots, 5 $, 
  and for other classical
gauge groups in \cite{KLEMM,ITO,RYANG,THEISEN}. In the 
particular
case of massless
$SU(2)$, the
 solutions to the PF equations are
given by hypergeometric functions \cite{BILAL,ITO}, while for 
$N_f=0$ 
$SU(3)$, they are given (in
certain regions of moduli space) by Appell functions, which
 generalise the hypergeometric function
\cite{KLEMM}. In other regions of the
$SU(3)$ moduli space, only double power-series solutions are 
available. Thus, even given the
hyperelliptic curve characterising the Coulomb phase of an $N=2$
 supersymmetric gauge theory,
considerable analysis is required to obtain the SW periods and the
 effective Lagrangian in various
regions of moduli space.

The first task in such a programme is to obtain the PF equations 
for the SW period integrals. Klemm
{\it et al\/.} \cite{KLEMM} describe a particular procedure which
 enables them to obtain the PF equations
for
$S
U(3)$ with $N_f=0$, which in principle is applicable to other
 theories as well. One would like to
obtain and solve the PF equations for a wide variety of theories 
in order to explore the physics
contained in particular solutions, and also to obtain an 
understanding of the general features of 
$N=2$ gauge theories. Therefore it is helpful to have an efficient 
method for constructing PF
equations from a given hyperelliptic curve, so that one can obtain
 explicit solutions for groups
with rank greater than or equal to 2.

It is the purpose of this paper to present a systematic method for 
finding the PF equations for the
SW periods which is particularly convenient for symbolic computer 
computations, once  the
hyperelliptic curve appropriate to a given $N=2$ supersymmetric
 gauge theory is known. Our method
should be considered as an alternative to that of Klemm 
{\it et al\/.} \cite{KLEMM}. A key element in our
treatment is  the Weyl group symmetry underlying the algebraic 
curve
 that
 describes the vacuum
structure of the effective $N=2$ SYM  theory (with or without 
massless hypermultiplets). For
technical reasons, we will not treat the theories with non-zero 
bare masses, but leave a discussion
of such cases to subsequent work \cite{NOS}.

This paper is organised as follows. In section 2, our method is 
described in general, so that given
a hyperelliptic curve for some $N=2$ theory, one will obtain a 
coupled set of partial, first-order
differential  equations for the periods. The method is further 
elucidated in section 3,
where a technique is developed to obtain a decoupled set of  partial,
 second-order differential
equations satisfied by the SW periods. A number of technical
details pertaining to the application of our method to  different 
gauge groups (both classical and
exceptional) are also given in section 3. Some relevant examples in
 rank 2 are worked out in
detail in section 4, for illustrative purposes. Our results are 
finally summarised in section 5.

Appendix A deals with a technical proof that is omitted from the 
body
 of the text. An extensive
catalogue of results is presented in  appendix B, including
 $N_f\neq 0$ theories (but always with zero bare mass).
 Explicit solutions to the PF equations themselves for rank greater
than 2 can be quite
 complicated, so  we will
restrict this paper to the presentation of the methods and a 
catalogue of PF equations. Solutions to some
interesting cases will be presented in a sequel in preparation
\cite{NOS}.  The methods of this paper
will have applications to a variety of questions, and are not 
limited to the SW problem.

\renewcommand{\theequation}{2.\arabic{equation}}
\setcounter{equation}{0}

\section{The Picard-Fuchs Equations: Generalities}

\subsection{Formulation of the problem}

Let us consider the complex algebraic curve
\be
y^2=p^2(x)-x^k\Lambda ^l
\label{eq:za}
\ee
where $p(x)$ is the polynomial
\be
p(x)=\sum_{i=0}^n u_i x^i=x^n+u_{n-2}x^{n-2}+\ldots+
u_1x+u_0
\label
{eq:zai}
\ee
$p(x)$ will be  the characteristic polynomial corresponding to 
  the fundamental representation of
the Lie algebra of the effective $N=2$ theory. We can therefore
 normalise the leading coefficient 
to 1, \ie,  $u_n=1$. We can  also take  $u_{n-1}=0$, as  all
 semisimple Lie algebras can be
represented by traceless matrices.  The integers $k$, $l$ and $n$,
 as
well as  the required coefficients $u_i$ corresponding to  various
 choices of a gauge group  and
matter content, have been
 determined in 
\cite{LERCHE}--\cite{DANI}.
 From dimensional analysis  we have
$0\leq k<2n$ \cite{SW}--\cite{MINAHAN}.
$\Lambda$ is the quantum scale of the effective $N=2$ theory.
 Without loss of generality, we will 
set $\Lambda=1$ for simplicity in what follows. If needed, the
 required powers of $\Lambda$ can be
reinstated by imposing the condition of homogeneity of the 
equations
 with respect to the (residual)
$R$-symmetry.

Equation \ref{eq:za} defines a family of hyperelliptic  Riemann 
surfaces
$\Sigma_g$ of genus
$g=n-1$  \cite{FARKAS}. The  moduli space of the curves \ref{eq:za} 
coincides
with the  moduli
space of quantum  vacua of the  $N=2$ theory
under consideration. The coefficients  $u_i$ are called the 
{\it moduli} of the surface. On
$\Sigma_g$ there are $g$ holomorphic 1-forms which, in the canonical
 representation, can be expressed as
\be
x^j\,{\d x\o y},\qquad j=0, 1, \ldots, g-1
\label{eq:zb}
\ee
and are also called {\it abelian differentials of the first 
kind}\/.
The following  $g$  1-forms  are meromorphic on $\Sigma_g$ and 
have 
vanishing residues:
\be
 x^j\,{\d x\o y},\qquad j=g+1, g+2, \ldots, 2g
\label{eq:zbi}
\ee
Due to this property of having  zero residues, they are  also called 
  {\it abelian differentials of
the second kind}\/. Furthermore, the 1-form
\be
x^g{\d x\o y}
\label{eq:zza}
\ee
is also meromorphic on $\Sigma_g$, but with non-zero residues. 
Due to
 this property of having non-zero
residues  it is also called an {\it abelian differential of the third
 kind}\/. Altogether, the  
abelian differentials $x^j\d x/y$ in equations \ref{eq:zb} and
\ref{eq:zbi}  will be denoted collectively by $\omega_j$, where 
$j=0, 1, \ldots, 2g$, $j\neq g$.   We define
the {\it basic range} $R$ to be $R=\{0, 1, \ldots, \check g,\ldots
 2g\}$, where a check over $g$
means the value $g$ is to be omitted. 

In effective 
$N=2$ supersymmetric gauge theories, there exists a preferred 
differential $\lambda_{SW}$, called
the Seiberg-Witten (SW) differential, with the following property 
\cite{SW}: the electric and magnetic
masses
$a_i$ and $a^D_i$ entering the BPS mass formula are given by the
 periods of $\lambda_{SW}$ along some
specified closed cycles $\gamma_i, \gamma^D_i \in H_1(\Sigma_g)$,
 \ie, 
\be
a_i=\oint_{\gamma_i} \lambda_{SW},\qquad a^D_i=\oint_
{\gamma^D_i} 
\lambda_{SW}
\label{eq:zc}
\ee
The SW differential further enjoys the property that its modular 
derivatives 
$\p\lambda_{SW}/\p u_i$ are (linear combinations of the) 
holomorphic
 1-forms \cite{SW}. This ensures
positivity of the K\"ahler metric on moduli space. Specifically, for 
the curve given in \ref{eq:za} we have
\cite{ARG,OZ,DANI}
\be
\lambda_{SW}=\Big[{k\o 2} p(x) - x p'(x)\Big]{\d x\o y}
\label{eq:ze}
\ee
In the presence of non-zero (bare) masses for matter
hypermultiplets, the SW differential picks up a non-zero residue
\cite{SW}, thus causing it to be of the third kind. Furthermore, when 
the
matter hypermultiplets are massive,  the SW differential is no longer
given by equation \ref{eq:ze}. In what follows $\lambda_{SW}$ 
will never be
of the third kind, as we are restricting ourselves to the pure SYM
theory, or to theories with  massless matter. 

Let us define $W=y^2$, so equation \ref{eq:za} will read
\be
W=p^2(x)-x^k=\sum_{i=0}^n\sum_{j=0}^n u_iu_j x^{i+j} -x^k
\label{eq:zf}
\ee
Given any differential $x^m\d x/y$, with $m\geq 0$   an integer, let
 us define its {\it
generalised $\mu$-period} $\Omega_m^{(\mu)}(u_i;\gamma)$ 
along a
 fixed 1-cycle $\gamma\in
H_1(\Sigma_g)$  as the line integral \cite{MUKHERJEE}
\be
\Omega_m^{(\mu)}(u_i; \gamma):=(-1)^{\mu +1}\Gamma 
(\mu + 1) 
\oint_{\gamma}{x^m\o W^{\mu + 1}}\,\d x
\label{eq:zg}
\ee
In equation \ref{eq:zg}, $\Gamma(\mu)$ stands for Euler's 
gamma function, 
while $\gamma\in  H_1(\Sigma_g)$ is
any  closed 1-cycle on the surface. As $\gamma$ will be  arbitrary 
but otherwise kept fixed,  
$\gamma$ will not appear explicitly in the notation. The {\it usual} 
periods of the Riemann
surface (up to an irrelevant normalisation factor) are of course 
obtained upon setting $\mu =-1/2$,  taking $m=0,1,\ldots, g-1$, 
and 
$\gamma$ to run over a canonical (symplectic) basis of
$H_1(\Sigma_g)$ \cite{FARKAS}. However, 
we will find it convenient to work with an arbitrary $\mu$ which 
will only be set equal to $-1/2$  at
the very end. 
The objects $\Omega_m^{(\mu)}$, and the differential
 equations they satisfy (called
Picard-Fuchs (PF) equations),  will be our prime focus of attention. 
With abuse of language, we will
continue to call the $\Omega_m^{(\mu)}$ {\it periods}, with the 
added
 adjectives {\it of the first,
second}, or {\it third kind},  if $m=0,
\ldots, g-1$, $m=g+1, \ldots, 2g$,  or  $m=g$, respectively.

\subsection{The recursion relations}

We now proceed to derive a set of recursion relations that will be
used to set up to PF equations.

{}From equation \ref{eq:zf} one easily finds
\begin{eqnarray}
{\p W\o\p x}&=&2 n x^{2n-1}+\sum_{j=0}^{n-1} \sum_{l=0}
^{n-1}(j+l)
u_ju_l x^{j+l-1}
+2\sum_{j=0}^{n-1}(n+j) u_j x^{n+j-1}-kx^{k-1}\nonumber\\
{\p W\o\p u_i}&=&2\sum_{j=0}^n u_j x^{i+j}
\label{eq:zh}
\end{eqnarray}
Solve for the highest power of $x$  in $\p W/\p x$, \ie, $x^{2n-1}$,
in equation
\ref{eq:zh},  and substitute the result into
\ref{eq:zg} to find
\begin{eqnarray}
\Omega_{m+2n-1}^{(\mu +1)} & = &
(-1)^{\mu +2}\Gamma(\mu +2)\oint_{\gamma} {x^m\o W^
{\mu +2}}
\Big[{1\o 2n}{\p W\o\p x}\nonumber \\
& - &{1\o 2n}\sum_{j=0}^{n-1}\sum_{l=0}^{n-1}(j+l)u_ju_lx^{j+l-1}-
{1\o
n}\sum_{j=0}^{n-1}(n+j) u_j x^{n+j-1}+{k\o 2n}x^{k-1}\Big]= 
\nonumber \\
& - &{m\o 2n} \Omega_{m-1}^{(\mu)}-{1\o
2n}\sum_{j=0}^{n-1}\sum_{l=0}^{n-1}(j+l)u_ju_l\Omega
_{m+j+l-1}^
{(\mu +1)}\nonumber \\
& - &{1\o n}\sum_{j=0}^{n-1}(n+j)u_j\Omega_{m+n+j-1}
^{(\mu+1)}
+{k\o 2n}\Omega_{m+k-1}^{(\mu + 1)}
\label{eq:zi}
\end{eqnarray}
To obtain the last line of equation \ref{eq:zi}, an integration 
by parts has
been
 performed   and a total
derivative dropped. If $m\neq 0$, shift $m$ by one unit to 
obtain 
from equation \ref{eq:zi}
\be
\Omega_m^{(\mu)}={1\o m+1}\Big[k\Omega_{m+k}
^{(\mu +1)}-2n \Omega_
{m+2n}^{(\mu +1)}
-\sum_{j=0}^{n-1}\sum_{l=0}^{n-1}(j+l)u_ju_l\Omega
_{m+j+l}^{(\mu +1)}
-2\sum_{j=0}^{n-1}(n+j)u_j\Omega_{m+n+j}^{(\mu +1)}
\Big]
\label{eq:zj}
\ee
Next we use equation \ref{eq:zf} to compute
\bea
& - & (1+\mu )\Omega_m^{(\mu)}=(-1)^{\mu +2}\Gamma 
(\mu+2) \oint_{\gamma}
\frac{x^m W}{W^{\mu+2}}  = \nonumber \\
& = &(-1)^{\mu +2}\Gamma(\mu+2)\oint_{\gamma}{x^m \o 
W^{\mu+2}}
\Big[\sum_{l=0}^n\sum_{j=0}^n u_lu_j x^{l+j} -x^k\Big]=
\nonumber \\
& = & \sum_{l=0}^n\sum_{j=0}^n u_lu_j\Omega_{m+l+j}
^{(\mu +1)}-\Omega_
{m+k}^{(\mu +1)}
\label{eq:zk}
\eea
and use this to solve for the period with the highest value of the 
lower index, $m+2n$, to get
\be
\Omega_{m+2n}^{(\mu+1)}=\Omega_{m+k}^{(\mu+1)}-
\sum_{j=0}^{n-1}\sum_
{l=0}^{n-1} u_ju_l 
\Omega_{m+j+l}^{(\mu+1)} -2\sum_{j=0}^{n-1} u_j 
\Omega_{m+n+j}
^{(\mu+1)} -(1+\mu)\Omega_m^{(\mu)}
\label{eq:zl}
\ee
Replace $\Omega_{m+2n}^{(\mu+1)}$ in equation \ref{eq:zj}
 with its value from
\ref{eq:zl}  to arrive at
\bea
\lefteqn{
\Omega_m^{(\mu)}=
{1\o m+1-2n(1+\mu)}\Big[(k-2n)\Omega_{m+k}^{(\mu+1)}}
\nonumber \\
&&+\sum_{j=0}^{n-1}\sum_{l=0}^{n-1}(2n-j-l)u_ju_l
\Omega_{m+j+l}^
{(\mu+1)}+
2\sum_{j=0}^{n-1}(n-j)u_j\Omega_{m+n+j}^{(\mu+1)}\Big]
\label{eq:zm}
\eea
Finally, take $\Omega_m^{(\mu)}$ as given in equation \ref{eq:zm} 
and
substitute it
 into \ref{eq:zl} to obtain an equation
involving $\mu +1$ on both sides. After shifting $m\rightarrow m-2n$,
 one gets
\bea
\Omega_{m}^{(\mu+1)} & = & {1\o m+1-2n(\mu+2)}\Big[\big(m-2n
+1-k (1+\mu)
\big)\Omega_{m+k-2n}^{(\mu+1)}\nonumber \\
&+ & 2\sum_{j=0}^{n-1}\big((1+\mu)(n+j)-(m-2n+1)\big)u_j
\Omega_{m-n+j}^
{(\mu+1)}\nonumber \\
&+ & \sum_{j=0}^{n-1}\sum_{l=0}^{n-1}\big((j+l)(1+\mu)-
(m-2n+1)\big)u_j
u_l\Omega_{m+j+l-2n}^{(\mu+1)}
\Big]
\label{eq:zmk}
\eea
We  now set $\mu=-1/2$ and
 collect the two recursion relations 
\ref{eq:zm} 
and \ref{eq:zmk}
\bea
\Omega_m^{(-1/2)} & = & {1\o m-(n-1)}\Big[(k-2n)
\Omega_{m+k}^
{(+1/2)}\nonumber \\
&+ & \sum_{j=0}^{n-1}\sum_{l=0}^{n-1}(2n-j-l)u_ju_l
\Omega_{m+j+l}^
{(+1/2)}+
2\sum_{j=0}^{n-1}(n-j)u_j\Omega_{m+n+j}^{(+1/2)}
\Big]
\label{eq:zn}
\eea
and
\bea
\Omega_{m}^{(+1/2)}& = & {1\o
m+1-3n}\Big[\big(m-2n+1-{k\o 2}\big)\Omega
_{m+k-2n}^{(+1/2)}\nonumber \\
&+ & \sum_{j=0}^{n-1}\big(n+j-2(m-2n+1)\big)u_j
\Omega_{m-n+j}^{(+1/2)
}
\nonumber \\
&+ & \sum_{j=0}^{n-1}\sum_{l=0}^{n-1}\big({1\o
2}(j+l)-(m-2n+1)\big)u_ju_l\Omega_{m+j+l-2n}^{(+1/2)}
\Big]
\label{eq:zo}
\eea

Let us now pause  to explain the significance of equations 
\ref{eq:zn} and 
\ref{eq:zo}. The existence of a
particular symmetry of the curve under consideration may simplify
 the  analysis of these equations.
For the sake of clarity, we will for the moment assume that equation 
\ref{eq:zn} will not have to be
evaluated at 
$m=g=n-1$, where it blows up
, and that its right-hand side will not 
contain occurrences of the
corresponding period $\Omega_{n-1}^{(+1/2)}$. A similar 
assumption 
will be made regarding equation
\ref{eq:zo}, \ie, it will not have to be evaluated at $m=3n-1$, and is
 right-hand side will not contain the
period $\Omega_{n-1}^{(+1/2)}$. In other words, we are for the 
moment assuming that we can
restrict ourselves to the subspace of differentials $\omega_j$ 
where $j\in R$ (or likewise for the corresponding
periods $\Omega_j
^{(\pm1/2)}$). This is the subspace of
differentials of the first and second kinds, \ie, the 1-forms with
vanishing residue. For a given curve, the particular 
subspace of differentials that one
has to restrict to depends on the corresponding gauge group; this 
will be explained in section 3, where these issues are dealt 
with. For the sake of the present discussion, the particular subspace
of differentials that we are
restricting to  only serves an illustrative purpose.

Under such assumptions, equation \ref{eq:zn} 
expresses $\Omega_m^{(-1/2)}$ as a linear combination, with 
$u_i$-dependent coefficients, of 
periods $\Omega_l^{(+1/2)}$. As $m$ runs over $R$, the linear
 combination in the
right-hand side of equation \ref{eq:zn} contains increasing values 
of $l$,
 which will eventually lie outside
$R$. We can bring them back into $R$ by means of equation 
\ref{eq:zo} it is
 a recursion relation expressing
$\Omega_l^{(+1/2)}$ as a linear combination (with $u_i$-dependent
 coefficients) of
$\Omega_l^{(+ 1/2)}$  with lower values of the subindex. Repeated
 application of equations \ref{eq:zn} and
\ref{eq:zo} will eventually allow one to express $\Omega_m^{(-1/2)}$, 
where
 $m\in R$, as a linear
combination of
$\Omega_l^{(+1/2)}$, with $l\in R$. The coefficients entering those 
linear combinations will be
some polynomials in the moduli $u_i$, in principle computable using 
the above recursion relations.
Let us call $M^{(-1/2)}$ the matrix of such coefficients
\cite{MUKHERJEE}
. Suppressing
 lower indices for
simplicity, we have 
\be
\Omega^{(-1/2)}= M^{(-1/2)}\cdot \Omega^{(+1/2)}
\label{eq:zp}
\ee
where the $\Omega_l^{(\pm 1/2)}$, $l\in R$, have been arranged 
as a 
column vector. We will from now
on omit the superscript $(-1/2)$ from $M^{(-1/2)}$, with the 
understanding that the value $\mu=-1/2$
has been fixed.

\subsection{Derivation of the Picard-Fuchs equations}

Having derived the necessary recursion relations, we can now
start taking modular derivatives of the periods.
{}From equation \ref{eq:zh} and the definition of the periods 
\ref{eq:zg} we have
\be
{\p \Omega_m^{(\mu)}\o\p u_i} =(-1)^{\mu+2}\Gamma
(\mu+2)\oint_{\gamma}
{x^m\o W^{\mu+2}}{\p W\o\p u_i}=
2\sum_{j=0}^n u_j \Omega_{m+i+j}^{(\mu+1)}
\label{eq:zq}
\ee
Again, the right-hand side of equation \ref{eq:zq} will eventually 
contain 
values of the lower index outside
the basic range $R$, but use of the recursion relations above will
 reduce it to a linear combination
of periods 
$\Omega_l^{(\mu+1)}$ with $l\in R$. The 
coefficients will 
be  polynomials in the moduli $u_i$; let
$D(u_i)$ be this matrix of coefficients. Setting $\mu=-1/2$, we end 
up with a system of equations
which, in matrix form, reads
\be
{\p\o\p u_i}\Omega^{(-1/2)}=D(u_i)\cdot \Omega^{(+1/2)}
\label{eq:zr}
\ee
As a second assumption to be justified presently, suppose for the 
moment that the matrix $M$ in
equation \ref{eq:zp} can be inverted  to solve for
$\Omega^{(+1/2)}$ as a function of $\Omega^{
(-1/2)}$. 
Substituting 
the result into equation \ref{eq:zr}, one gets
\be
{\p\o\p u_i}\Omega^{(-1/2)}=D(u_i)\cdot M^{-1}\cdot
\Omega^{(-1/2)}
\label{eq:zs}
\ee
Equation \ref{eq:zs} is a coupled system of first-order, partial
 differential equations for the periods
$\Omega^{(-1/2)}$. The  coefficients are  rational functions 
of the
 moduli $u_i$, computable from a
knowledge of $W$ and  the recursion relations derived above. In 
principle, integration of this
system of equations yields the periods as functions of the moduli
 $u_i$. The particular 1-cycle
$\gamma\in H_1(\Sigma_g)$  being integrated over appears 
in the specific choice  of 
boundary conditions that one  makes. In practice, however, the
 fact  that
the system \ref{eq:zs} is coupled makes it very difficult to solve. A
 possible strategy is to concentrate
on one particular period and try to obtain a reduced system of 
equations  satisfied by it.
Decoupling of the equations may be achieved at the cost of 
increasing the order of derivatives. Of
course, in the framework of effective $N=2$ SYM  theories,
 one is especially interested in
obtaining a system of equations satisfied by the periods of the SW 
differential $\lambda_{SW}$.

In what follows we will therefore concentrate on solving the problem 
for the SW periods within the subspace of differentials
with vanishing residue, as assumed in section 2.2. In order to do so,
the first step is to include the differential $\lambda_{SW}$  as a
basis vector by means of 
a change of basis. From equations 
\ref{eq:zai} and
\ref{eq:ze} we have
\be
\lambda_{SW}=\sum_{j=0}^n\big({k\o 2}-j\big)u_jx^j\,
{\d x\o y}
\label{eq:zt}
\ee
We observe that $\lambda_{SW}$ is never of the third kind, 
because
$u_g=u_{n-1}=0$.   As $k<2n$ and $u_n=1$, $\lambda_{SW}$ 
always
carries a nonzero  component along $x^{g+1}\d x/y$, so we
can take the new basis of differentials of the first and second
kinds to be spanned by 
\be
x^i\, {\d x\o y}, \quad {\rm for} \;i\in\{0, 1, \ldots,
 g-1\}, 
\quad\lambda_{SW}, \quad x^j \,{\d x\o
y},\quad {\rm for}\;j\in\{g+2,\ldots, 2g\}
\label{eq:zu}
\ee
We will find it convenient to arrange the new basic differentials 
in this order. Call  $K$  the
matrix implementing this change of basis from the original one in 
equations \ref{eq:zb} and \ref{eq:zbi} to the above in equation 
\ref{eq:zu}; one easily
checks that ${\rm det}\, K\neq 0$. If $\omega$ and $\pi$  are 
column
vectors  representing the
old and new basic differentials, respectively, then from the matrix 
expression
\be
K\cdot\omega =\pi
\label{eq:zv}
\ee
there follows a similar relation for the corresponding periods,
\be
K\cdot\Omega =\Pi
\label{eq:zw}
\ee
where $\Pi$ denotes the periods associated with the new basic 
differentials, \ie, those defined in equation \ref{eq:zu}.  Converting 
equation \ref{eq:zs} to the new basis is immediate:
\be
{\p\o\p u_i}\Pi^{(-1/2)}=\Big[K\cdot D(u_i)\cdot M^{-1}
\cdot K^{-1} 
+{\p K\o\p u_i}\cdot
K^{-1}
\Big]\, \Pi^{(-1/2)}
\label{eq:zx}
\ee
Finally, define $U_i$ to be 
\be
U_i:=\Big[K\cdot D(u_i)\cdot M^{-1}\cdot K^{-1} +
{\p K\o\p u_i}
\cdot K^{-1}\Big]
\label{eq:zy}
\ee
in order to reexpress equation \ref{eq:zx} as
\be
{\p\o\p u_i}\Pi^{(-1/2)}=U_i\cdot \Pi^{(-1/2)}
\label{eq:zz}
\ee
The matrix $U_i$ is computable from the above; its entries are 
rational functions of the moduli
$u_i$.

The invertibility of $M$ remains to be addressed. Clearly, as the 
definition of the $M$ matrix
requires restriction to 
an appropriate subspace of differentials, 
this issue will have to be dealt
with on a case-by-case basis. However, some general arguments 
can be 
put forward. From \cite{FULTON} we
have the following decomposition for the discriminant $\Delta(u_i)$ 
of the curve:
\be
\Delta (u_i)=a(x) W(x) + b(x) {\p W(x)\o\p x}
\label{eq:zzi}
\ee
where $a(x)$ and $b(x)$ are certain polynomials in $x$. This 
property 
is used in
\cite{KLEMM} as follows. Taking  the modular derivative 
$\p/\p u_i$ of 
the pe
riod integral causes the
power in the denominator to increase by one unit, 
as in equation \ref{eq:zq} 
$\mu+1\rightarrow \mu+2$. In \cite{KLEMM},
this exponent is made to decrease again by use of the formula
\be
{\phi(x)\o W^{\mu/2}}={1\o \Delta(u_i)}{1\o W^{\mu/2-1}}
\Big(a\phi+{2\o\mu-2}{\d\o\d x}(b\phi)\Big)
\label{eq:zzii}
\ee
where $\phi(x)$ is any polynomial in $x$. Equation \ref{eq:zzii} is
valid only under the integral sign. It ceases to hold when the curve
 is singular, \ie,
at those 
points of moduli space such that  $\Delta(u_i) =0$. The
 defining equation of the
$M$ matrix, \ref{eq:zp}, is equivalent to  equation \ref{eq:zzii}, 
when  the latter 
is read from right to left, \ie,
in decreasing order of $\mu$. We therefore expect $M$ to be
invertible
 except at the singularities
of moduli space, \ie, on the zero locus of $\Delta(u_i)$. A  
proof of this fact is given in
appendix A.

To further elaborate on the above  argument, let us observe that the
 homology cycles of
$H
_1(\Sigma_g)$ are defined so as to encircle the zeroes of $W$. 
A vanishing discriminant
$\Delta(u_i)=0$ at some given point of moduli space implies the 
vanishing of the homology cycle that
encircles the two collapsing roots,
\ie, a degeneration of $\Sigma_g$. With this vanishing cycle there 
is also some differential in the cohomology of $\Sigma_g$ 
disappearing  as well. We therefore expect the PF
equations to exhibit some type of singular behaviour when 
$\Delta(u_i)=0$, as they in fact do.

Equation \ref{eq:zz} is the most general expression that one can derive
 without making any specific
assumption as to the nature of the gauge group or the (massless) 
matter content of the theory. From
now on, however, a case-by-case analysis is necessary, as  required
 by the different gauge groups.
This is natural, since the SW differential depends on the choice of 
a gauge group and matter
content. However, some general features do emerge, which allow 
one to observe a general pattern, as
will be explained in the following section.

\section{Decoupling the Picard-Fuchs Equations}

\subsection{The $B_r$ and $D_r$ gauge groups}

\renewcommand{\theequation}{3.\arabic{equation}}
\setcounter{equation}{0}

Let us first  consider the $SO(2r+1)$ and $SO(2r)$ gauge
 theories\footnote{Although there exists a well defined relation 
between
the rank
$r$ and the genus $g=n-1$ of the corresponding curve, we will not
require it, and will continue to use $n$, rather than its expression
as a function of $r$. For the gauge groups in this section, we have
$g=2r-1$.}, either for the pure SYM case, or in the presence of
massless matter hypermultiplets in
 the fundamental representation.
We restrict ourselves to asymptotically free theories.  From
\cite{BRAND,DANI,DANII}, the polynomial $p(x)$ of equation 
\ref{eq:zai} is
even, as
$u_{2j+1}=0$. Therefore, the  curves
\ref{eq:za} describing moduli space are  invariant under an 
$x\rightarrow -x$ symmetry.
 This invariance is a consequence of two facts: a 
${\bf Z}_2$ factor present in the
corresponding Weyl groups, which causes the odd Casimir operators
 of the group to vanish, and the
property that the Dynkin index of the fundamental representation is
 even.  This symmetry turns out
to be useful in decoupling the PF equations, as it determines the 
right subspace of differentials
that one must restrict to.

Call a differential $\omega_m=x^m\d x /y$  {\it even} (respectively,
  {\it odd}\/) if 
$m$ is even (respectively, odd).\footnote{This definition excludes the 
$\d x$ piece of the differential;
thus, {\it e.g},  $x\,\d x/y$ is defined to be odd. Obviously, this
is  purely a matter of convention.}  We will  thus talk about  even or
odd periods accordingly. 
{}From equation \ref{eq:zt} we have that $\lambda_{SW}$ is even 
for these
gauge groups.  One also sees  from equations \ref{eq:zn} and 
\ref{eq:zo} that the 
recursion relations involved in
deriving the matrices $D(u_i)$ and $M$ do not mix   even with  odd 
periods, as they always have a step of
two units. This is a natural decoupling which strongly suggests 
omitting the odd and  restricting to
the even periods, something we henceforth do.  In particular, the 
matrix $M$ of equation \ref{eq:zp} will
also be restricted to this even subspace; we will check that 
${\rm det}\, M$ then turns out to be
proportional to (some  powers of) the factors of  $\Delta(u_i)$. 
The genus $g=n-1$ is always
odd, so the periods $\Omega_{n-1}^{(\pm1/2)}$
do not appear after such  a restriction. Another
consequence is that   the values $m=n-1$ and $m=3n-1$ at which 
equations \ref{eq:zn} and \ref{eq:zo} blow
up  are automatically jumped over by the recursions. 

The  even basic  differentials of equation \ref{eq:zu} are
\be
{\d x\o y}, \quad x^2{\d x\o y},\quad \ldots, x^{g-1}
{\d x\o y},
\quad \lambda_{SW},\quad x^{g+3}{\d x\o
y},\ldots, x^{2g}{\d x\o y}
\label{eq:zaai}
\ee
and  there are $n$ of them. We have that $n$ itself is even,  \ie,
the  subspace of even differentials  is
even-dimensional, so $n=2s$ for some $s$.\footnote{The precise 
value of
$s$ can be given as a function of the gauge group, \ie, as a
function of $n$, but it is irrelevant to the present discussion.} 
According to the notation
 introduced in equation \ref{eq:zv}, let
us denote the  basic differentials  of equation \ref{eq:zaai} by
\be
\{\pi_1, \ldots, \pi_{s}, \pi_{s+1}=\lambda_{SW}, 
\pi_{s+2}, 
\ldots,
\pi_{2s}\}
\label{eq:zaaii}
\ee
where, for the sake of clarity, indices have been relabelled so as 
to run from 1 to $2s$.
In equation \ref{eq:zaaii}, all differentials preceding  
$\lambda_{SW}=\pi_
{s+1}$ are of
the first kind; from
$\lambda_{SW}$ onward, all differentials are of the second kind. 
The  periods corresponding to the
differentials of equation \ref{eq:zaaii} are
\be
\{\Pi_1, \ldots, \Pi_{s}, \Pi_{s+1}=\Pi_{SW}, \Pi_{s+2}, 
\ldots,\Pi_
{2s}\}
\label{eq:zabi}
\ee
Notice that this restriction to the even
subspace is compatible with the change of 
basis implemented by 
$K$,
 \ie, $K$ itself did not mix even with odd differentials.

Once restricted to the even subspace, equation \ref{eq:zz}  reads
\be
{\p\o\p u_i}
\left( \begin{array}{c}
\Pi_1\\
\vdots\\
\Pi_{s}\\
\Pi_{s+1}\\
\vdots\\
\Pi_{2s}
\end{array} \right)
=
\left( \begin{array}{cccccc}
U^{(i)}_{11}&\ldots&U^{(i)}_{1s}&U^{(i)}_{1 s+1}&\ldots&
U^{(i)}_{1 2s}
\\
\vdots&\ddots&\vdots&\vdots&\ddots&\vdots\\
U^{(i)}_{s1}&\ldots&U^{(i)}_{ss}&U^{(i)}_{s s+1}&\ldots&
U^{(i)}_
{s
 2s}\\
0&1&0&0&\ldots&0\\
\vdots&\ddots&\vdots&\vdots&\ddots&\vdots\\
U^{(i)}_{2s1}&\ldots&U^{(i)}_{2ss}&U^{(i)}_{2s s+1}&
\ldots&U^{(i)}_
{2s 2s} 
\end{array}
\right)
\left( \begin{array}{c}
\Pi_1\\
\vdots\\
\Pi_s\\
\Pi_{s+1}\\
\vdots\\
\Pi_{2s}
\end{array}
\right)
\label{eq:zaa}
\ee
where the $(s+1)$-th row is everywhere zero, except at the $i$-th 
position, $1\leq i\leq s$,
whose entry is 1, so that
\be
{\p\o\p u_i}\Pi_{s+1}={\p\o\p u_i}\Pi_{SW}=\Pi_i,
\qquad 1\leq i\leq s
\label
{eq:zab}
\ee
Equation \ref{eq:zab} follows from  the property that the SW differential 
$\lambda_{SW}=\pi_{s+1}$ is a {\it
potential} for the even holomorphic differentials, \ie,  
\be
{\p\lambda_{SW}\o\p u_i}={\p\pi_{s+1}\o\p u_i}=\pi_i,\qquad
1\leq i
\leq s
\label{eq:zac}
\ee
since integration of equation \ref{eq:zac} along some 1-cycle  produces 
the  corresponding
statement for the periods. An analogous property for the odd periods 
does not hold, as all the odd
moduli $u_{2i+1}$ vanish by symmetry.

To proceed further, consider the  $U_i$ matrix in equation \ref{eq:zaa} 
and block-decompose it as 
\be
U_i=\pmatrix{A_i&B_i\cr C_i&D_i}
\label{eq:zacc}
\ee
where all four blocks $A_i$, $B_i$, $C_i$ and $D_i$ are $s\times s$. 
Next
take the equations for the derivatives of holomorphic periods $\p
\Pi_j/\p u_i$ ,  $1\leq j\leq s$,
and solve them for the meromorphic periods $\Pi_j$, $s\leq j\leq 2s$,
 in terms of the holomorphic
ones and their modular derivatives. That is, consider

\be
{\p\o\p u_i}\pmatrix{\Pi_1\cr\vdots\cr\Pi_s\cr}- A_i\pmatrix{\Pi_1
\cr\vdots\cr\Pi_s\cr}=
B_i\pmatrix{\Pi_{s+1}\cr\vdots\cr\Pi_{2s}\cr}
\label{eq:zad}
\ee
Solving equation \ref{eq:zad} for the meromorphic periods involves 
inverting
the  matrix $B_i$. Although we lack a
formal proof that $B_i$ is invertible, ${\rm det}\, B_i$ turns out
 to  vanish on the zero locus
of  the discriminant of the curve in all the cases catalogued in
appendix B,  so $B_i$ will be invertible
except at the singularities of moduli space.  From equation \ref{eq:zad},
\be
\pmatrix{\Pi_{s+1}\cr\vdots\cr\Pi_{2s}\cr}=
B_i^{-1}\cdot\Big({\p\o\p u_i}-A_i\Big)\pmatrix{\Pi_1\cr\vdots\cr
\Pi_s\cr}
\label{eq:zae}
\ee
We are interested in the SW period $\Pi_{s+1}$ only, so we discard
 all  equations in \ref{eq:zae} but the
first one:
\be
\Pi_{s+1}=(B_i^{-1})_1^r \,{\p\Pi_r\o\p u_i} - (B_i^{-1}A_i)_1^r \,
\Pi_r
\label{eq:zaf}
\ee
where a sum over $r$, $1\leq r\leq s$, is implicit in  equation  
\ref{eq:zaf}. Finally, as the right-hand
side of equation 
\ref{eq:zaf} involves nothing but holomorphic periods and modular 
derivatives thereof,  we can use \ref{eq:zab} to
obtain an equation involving  the SW period $\Pi_{s+1}=\Pi_{SW}$ 
on both sides:
\be
\Pi_{SW}=(B_i^{-1})_1^r \,{\p^2\Pi_{SW}\o\p u_i\p u_r} -
 (B_i^{-1}
A_i)_1^r \,{\p\Pi_{SW}\o\p u_r}
\label{eq:zag}
\ee
Equation \ref{eq:zag} is a   partial differential equation, second-order
 in modular derivatives, which is
completely decoupled, \ie, it
 involves nothing but the SW period
 $\Pi_{SW}$. The number of such
equations equals the number of moduli; thus giving a decoupled 
system of partial, second-order
differential equations satisfied by the SW period: the desired PF
equations for the $N=2$ theory.

\subsection{The $C_{r}$ gauge groups}

Let us consider the $N_f>0$ $Sp(2r)$ gauge theory as described by the
curves given in \cite{EGUCHI,SASAKURA}.\footnote{The pure 
$Sp(2r)$ SYM
theory can be described by a curve whose polynomial $p(x)$ is even
\cite{IRAN}, so it can be studied by the methods of section
3.1}${^,}$\footnote{For $Sp(2r)$, we have $g=2r$.}
 As the Weyl group of $Sp(2r)$ contains a ${\bf
Z}_2$ factor, all the odd Casimir operators vanish, \ie, we have
$u_{2j+1}=0$. However, a close examination reveals that the curves
given in \cite{EGUCHI,SASAKURA} contain a factor of $x^2$ in the
left-hand side. Pulling this factor to the right-hand side has the
effect of causing the resulting polynomial $p(x)$ to be {\it odd}
 under $x\rightarrow -x$. The genus $g=n-1$ will
now be  even because the degree $n$ of this resulting $p(x)$ will  be
odd. The  ${\bf Z}_2$ symmetry dictated by the Weyl group is {\it not}
violated, as the complete curve $y^2=p^2(x)-x^k\Lambda ^l$ continues
to be even, since $k=2(N_f-1)$  is also even \cite{EGUCHI}.

From this one can  suspect that the right subspace of differentials
(or periods) that one must restrict to is given by the  odd
differentials of equation \ref{eq:zu}. That this is so is
 further confirmed
by the fact that the recursion relations  \ref{eq:zn} and \ref{eq:zo} now 
have a
step of 2 units, and that the SW differential $\lambda_{SW}$ will now
be odd, as  revealed by equations \ref{eq:ze} and \ref{eq:zt}. 
In consequence, the values $m=n-1$ and
$m=3n-1$ at which the recursions \ref{eq:zn} and \ref{eq:zo} blow 
up are jumped
over, and the periods $\Omega_{n-1}^{(\pm1/2)}$ do not appear.
As was the case for the $SO(2r)$
and $SO(2r+1)$ groups, this subspace of odd differentials is 
even dimensional. Furthermore,  the change of basis implemented by
the matrix $K$ of equation \ref{eq:zv} respects this even-odd 
partition, since $\lambda_{SW}$ is odd.

One technical point that appears for $Sp(2r)$, but not for the
orthogonal gauge groups, is the following. Let us remember that
$m=2g=2n-2$ is the highest value of $m$ such  that $m\in R$. We 
would
therefore expect the period $\Omega_{2n-1}^{(+1/2)}$ to be
expressible in terms
 of some $\Omega_{m}^{(+1/2)}$ with lower
 values of $m$, 
according to
equation \ref{eq:zo}. However, we
 cannot use equation \ref{eq:zo} to obtain this linear combination, 
since
the derivation of the latter relation  formally involved division by
zero when one takes $m=0$.  \footnote{Remember that $m$ was 
supposed to
be non-zero in passing from equation
\ref{eq:zi} to \ref{eq:zj}.}  Instead, we must return to equation \ref{eq:zi}  
and set
 $m=0$ to arrive at 
\be
\Omega_{2n-1}^{(+1/2)}=-{1\o
2n}\sum_{j=0}^{n-1}\sum_{l=0}^{n-1}(j+
l)u_ju_l\Omega_{j+l-1}
^{(+1/2)}
 -{1\o n}\sum_{j=0}^{n-1}(n+j)u_j\Omega_{n+j-1}^{(+1/2)} +{k\o
2n}\Omega_ {k-1}^{(+1/2)}
\label{eq:zcai}
\ee
As $2n-1$ is  odd, this period was omitted from the computations
of  the previous section, but it will be required for the resolution
of the recursion relations  for $Sp(2r)$.

With this proviso,  the same arguments explained for the 
$SO(2r)$
and $SO(2r+1)$ groups in   section  3.1 hold throughout, with the
only   difference that we will be working in the  subspace 
of odd  periods of the first and second kinds. As a consequence, the
SW differential
$\lambda_{SW}$ will be a potential for the odd differentials only.

\subsection{The $A_{r}$ gauge groups}

The Weyl group of $SU(r+1)$ does not possess a ${\bf Z}_2$ factor for
 $r>1$.\footnote{Obviously, $SU(2)$
is an exception to this discussion. The corresponding PF equations 
are very easy to derive and to
decouple for the SW period. See, \eg, 
\cite{BILAL,FERRARI,ITO,RYANG}.}{$^,$}\footnote{For 
$SU(r+1)$ we have $r=g$.} 
This implies the existence of
even as well as odd  Casimir operators for the group. Correspondingly,
 the  characteristic polynomial
$p(x)$ of  equation \ref{eq:zai} will also have non-zero odd moduli 
$u_{2j+1}$. In general, the SW differential will be neither even nor
odd, as  equation \ref{eq:ze} reveals. The same will be true for the
polynomial $p(x)$. The $x
\rightarrow -x$ symmetry used in the previous sections to decouple 
the PF equations, be it under its even or under its odd
presentation, is spoiled.

The first technical consequence of the above is that equation 
\ref{eq:zcai}
will have to be taken into consideration when solving the recursion
relations
\ref{eq:zn} and \ref{eq:zo}, because the period 
$\Omega_{2n-1}^{(+1/2)}$ will be
required.  Moreover, we have learned that  an essential  point  to be 
addressed is the identification of the appropriate subspace of
differentials (or periods) that we must
 restrict to.  It turns out
that the recursion relation \ref{eq:zo} must eventually be evaluated 
at 
$m=3n-1$. To prove this assertion,
consider the value $m=2n-2$ in equation \ref{eq:zn}, which is  
allowed since
 we still have $2g=2n-2\in R$. From
the right-hand side of this equation we find that, whenever 
$j+l=n+1$, the period
$\Omega_{3n-1}^{(+1/2)}$ is required. However, equation 
\ref{eq:zo} 
blows up when $m=3n-1$. We have seen that this
problem did not occur for the orthogonal and symplectic gauge 
groups.

The origin of this difficulty can be traced  back to the fact that, 
in the sequence of differentials
of the first and second kind  given in equations \ref{eq:zb} and 
\ref{eq:zbi}, 
there is a gap at $m=g$, since $x^g
\d x/y$ is always a differential of the third kind. As the recursion
 relations now have a step of one
unit, we cannot jump over the  value  $m=g=n-1$. To clarify this 
point, let us give an
alternative expression for $\Omega_{2n-2}^{(-1/2)}$ that will bear 
this out.

Consider $x^{2g}W=x^{2n-2} W$ as a
 polynomial in  $x$, and 
divide  
it by $\p W/\p x$ to obtain a
certain quotient $q(x)$, plus a certain remainder $r(x)$: 
\be
x^{2n-2}W(x)=q(x) {\p W(x)\o\p x} + r(x)
\label{eq:zca}
\ee
The coefficients of both $q(x)$ and $r(x)$ will be certain 
polynomial functions in the moduli
$u_i$, explicitly obtainable from \ref{eq:zca}.   The degree of 
$r(x)$ in 
$x$ will be $2n-2$, while that of
$q(x)$ will be $2n-1$, so let us put
\be
q(x)=\sum_{j=0}^{2n-1} q_j(u_i)x^j, \qquad r(x)=
\sum_{l=0}^{2
n-2}
r_l(u_i)x^l
\label{eq:zcb}
\ee
We have
\be
W(x)=x^{2n}+\ldots, \qquad {\p W(x)\o \p x} =2nx^{2n-1}
+\ldots, 
\qquad x^{2n-2}W(x)=x^{4n-2}+\ldots
\label{eq:zcbi}
\ee
so $q(x)$ must be of the form
\be
q(x)={1\o 2n} x^{2n-1} +\ldots
\label{eq:zcci}
\ee
Furthermore, from equation \ref{eq:zca}, 
\bea
&-&(1+\mu)\Omega_{2n-2}^{(\mu)}=(-1)^{\mu +2}
\Gamma(\mu+2)\oint_
{\gamma}{x^{2n-2} W\o W^{\mu+2}}=\nonumber \\ 
&=&(-1)^{\mu+2}\Gamma(\mu+2)\oint_{\gamma}
{1\o W^{\mu+2}}
\Big[\sum_{j=0}
^{2n-1}q_j(u_i)x^j{\p W\o\p x}+
\sum_{l=0}^{2n-2}r_l
(u_i)x^l\Big]
\label{eq:zcc}
\eea
Integrate by parts in the first summand of equation \ref{eq:zcc}
 to obtain
\be
-(1+\mu)\Omega_{2n-2}^{(\mu)}=-\sum_{j=0}^{2n-1}jq_j(u_i)
\Omega
_{j-1}^{(\mu)}+
\sum_{l=0}^{2n-2}r_l(u_i)\Omega_l^{(\mu+1)}
\label{eq:zcd}
\ee 
Now set $\mu=-1/2$ in  equation \ref{eq:zcd} and solve for
$\Omega_{2n-2}^{(-1/2)}$  using \ref{eq:zcci}:
\be
\Omega_{2n-2}^{(-1/2)}={2n\o
n-1}\Big(\sum_{l=0}^{2n-2}r_l\Omega_l^{(+1/
2)}-
\sum_{j=0}^{2n-2}jq_j
\Omega_{j-1}^{(-1/2)}\Big)
\label{eq:zce}
\ee 
Clearly, the right-hand side of equation \ref{eq:zce} will in 
general 
involve the periods
$\Omega_{n-1}^{(\pm 1/2)}$ corresponding to the gap 
in the sequence
 that defines the basic range
$R$. In principle, this implies that the subspace of differentials 
we must
 restrict to is that of the $\omega_m$ with 
$m\in R\cup\{n-1\}$.

Let us recall that $M$ is the matrix of coefficients in the 
expansion
 of $\Omega_m
^{(- 1/2)}$ as linear functions of the 
$\Omega_m^{(+ 1/2)}$.
 Inclusion of $\Omega_{n-1}^{(\pm 1/2)}$ would, in principle, 
increase the number of rows and columns of
$M$ by one unit, the increase being due to the expansion of
 $\Omega_
{n-1}^{(- 1/2)}$ as a
linear combination of the $\Omega_m^{(+ 1/2)}$, where 
$m\in R\cup
\{n-1\}$. However, we have no such
expansion at hand. We cannot define $M$ as  a $(2g+1)
\times(2g+1)$ 
matrix; the best we
can have is $2g$ rows (corresponding  to the $2g
$ periods 
$\Omega_m^
{(-1/2)}$, where $m\in R$), and
$2g+1$ columns (corresponding to the $2g+1$ periods 
$\Omega
_m^{(+1/2)}$, where $m\in R\cup
\{n-1\}$). As a non-square matrix cannot be invertible, 
this seems
 to imply the need to restrict
ourselves to a $2g\times 2g$ submatrix with maximal rank, 
and look 
for an invertible
$M$ matrix on that subspace. The procedure outlined in 
what follows 
serves precisely that 
purpose:

$\bullet$ Use equations \ref{eq:zn} and \ref{eq:zo} to express 
$\Omega_m^{(-1/2)}$,
 where $m\neq n-1$, as linear
combinations of the $\Omega_m^{(+1/2)}$, where $m\in 
R\cup\{n-1\}$, 
plus possibly also of
$\Omega_{n-1}^{(-1/2)}$. That the latter period can appear 
in the
 right-hand side of these
expansions has already been illustrated in equation \ref{eq:zce}.
 This gives 
us a $2g\times (2g+1)$ matrix.

$\bullet$ Any occurrence of $\Omega_{n-1}^{(- 1/2)}$ in the
 expansions
 that define the $2g$ rows
of $M$ is to be transferred to the left-hand 
side of the equations.
 Such occurrences
will only happen when $m>n-1$ as, for $m<n-1$,  equations 
\ref{eq:zn} and 
\ref{eq:zo} do not involve
$\Omega_{n-1}^{(\pm 1/2)}$. Transferring  
$\Omega_{n-1}^{(- 1/2)}$ to 
the left will also affect the
$D(u_i)$ matrices of equation \ref{eq:zr}: whenever the left-hand side 
presents occurrences of
$\Omega_{n-1}^{(-1/2)}$ (with $u_i$-dependent coefficients),  the
 corresponding  modular derivatives
will have to be modified accordingly.

$\bullet$ As the number of columns of $M$ will exceed that of rows
 by one,  a
linear dependence relation between the $\Omega_m^{(+ 1/2)}$ is 
needed. This will have the
consequence  of  effectively reducing  $M$ to a {\it square}  matrix.
 Only so will it
have a chance of being invertible, as required by the preceding 
sections.

In what follows we will derive the sought-for  linear dependence 
relation between the $\Omega_m^{(+
1/2)}$, where $m\in R \cup \{n-1\}$. The procedure is 
completely 
analogous to 
that
used in equations \ref{eq:zca} to \ref{eq:zce}. Consider
$x^{n-1}W$ as a polynomial in $x$, and divide it by $\p W/\p x$ 
to 
obtain a certain quotient
$\tilde q(x)$, plus a certain remainder $\tilde r(x)$, whose  
degrees are $n$ and $2n-2$,
respectively:
\be
\tilde q(x)=\sum_{j=0}^n \tilde q_j(u_i)x^j, \qquad \tilde r(x)=
\sum_{l=0}^{2n-2}\tilde r_l(u_i)x^l
\label{eq:zcf}
\ee
By the same arguments as in equations \ref{eq:zcbi} and 
\ref{eq:zcci}, we can write 
\be
\tilde q(x)={1\o 2n}
 x^n +\ldots
\label{eq:zcgg}
\ee
Furthermore, following the same reasoning as in equations 
\ref{eq:zcc} and 
\ref{eq:zcd}, we find

\bea
&-&(1+\mu)\Omega_{n-1}^{(\mu)}=(-1)^{\mu +2}
\Gamma(\mu+2)
\oint_{\gamma}
{x^{n-1} W\o W^{\mu+2}}=\nonumber \\
&=&-\sum_{j=0}^n j\tilde q_j(u_i)\Omega_{j-1}^
{(\mu)}+
\sum_{l=0}^{2n-2}\tilde r_l(u_i)\Omega_l^{(\mu+1)}
\label{eq:zch}
\eea
Setting $\mu=-1/2$ and solving equation \ref{eq:zch} for
 $\Omega_{n-1}^
{(-1/2)}$ produces
\be
0=\Big(
n\tilde q_n -{1\o 2}\Big) \Omega_{n-1}^{(-1/2)}=
-\sum_{j=0}^
{n-1} j\tilde
q_j(u_i)\Omega_{j-1}^{(-1/2)}+
\sum_{l=0}^{2n-2}\tilde r_l(u_i)\Omega_l^{(+1/2)}
\label{eq:zci}
\ee
where equation \ref{eq:zcgg} has been used to equate the 
left-hand side to
 zero. We therefore have
\be
\sum_{j=0}^{n-1} j\tilde q_j(u_i)\Omega_{j-1}^{(-1/2)}=
\sum_{l=0}^{2n-2}\tilde r_l(u_i)\Omega_l^{(+1/2)}
\label{eq:zcj}
\ee
Equation \ref{eq:zcj} is a linear dependence relation between 
$\Omega^
{(-1/2)}$ and $\Omega^{(+1/2)}$ which
does not involve $\Omega_{n-1}^{(-1/2)}$. Therefore, we 
are now able
 to  make use of equations \ref{eq:zn}
and
\ref{eq:zo}, \ie, of the allowed rows of $M$,  to recast the 
left-hand side
 of equation \ref{eq:zcj} as a linear
combination  of the $\Omega^{(+1/2)}$:
\be
\sum_{j=0}^{n-1}j\tilde q_j(u_i)\sum_{r\in R}
\big[M\big]_{j-1}^r 
\Omega_{r}^{(+1/2)}=
\sum_{l=0}^{2n-2}\tilde r_l(u_i)\Omega_l^{(+1/2)}
\label{eq:zcg}
\ee
Equation \ref{eq:zcg} is the sought-for linear 
dependence 
relation that 
appears due to the inclusion of 
$\Omega_{n-1}^{(\pm 1/2)}$. Restriction to the subspace 
defined by 
this relation produces a
{\it square}\/, $2g\times 2g$ matrix:
\be
\tilde\Omega^{(-1/2)}=  \tilde M \tilde\Omega
^{(+1/2)}
\label{eq:zchh}
\ee
The tildes in the notation remind us that the left-hand 
side will  
include 
occurrences of $\Omega_{n-1}^{(-1/2)}$ when $m>n-1$, 
possibly 
multiplied by some $u_i$-dependent
coefficients, while the right-hand side has been reduced as
 dictated
 by the linear dependence
relation \ref{eq:zcg}. {\it We claim that the $\tilde M$ matrix 
so defined
 is invertible, its
determinant vanishing on the zero locus of the discriminant
 $\Delta(u_i)$ of
the curve}\/. It is on this $2g$-dimensional space of differentials
 (or periods) that we will be
working. 

Let us make some technical observations on the procedure just 
described. In practice, restriction to
the subspace determined by  equation \ref{eq:zcg} means solving
for some given  $\Omega_{m}^{(+ 1/2)}$,
where $m\in R\cup\{n-1\}$,  as a linear combination 
(with $u_i$-dependent coefficients) of the
rest. The particular  $\Omega_m^{(+1/2)}$ that can be solved 
for  depends on the 
coefficients entering  equation \ref{eq:zcg}; any period whose 
coefficient 
is non-zero will do. Obviously, 
the particular
$\Omega_m^{(+1/2)}$ that is being solved for in equation 
\ref{eq:zcg} is 
irrelevant (as long as its
coefficient is non-zero), since any such $\Omega_
m^{(+1/2)}$ 
so 
obtained is just a different, but
equivalent,  expression of the same linear relation \ref{eq:zcg}.  
Whatever 
the choice, ${\rm det}\, \tilde
M$ will continue  to vanish on the zero locus of $\Delta(u_i)$.

However, differences may arise in the actual entries of $\tilde M$,
 due to the fact that
different (though equivalent) sets of basic $\Omega_m^{(+1/2)}$ 
are  being used. Once a given set of
$2g$ independent $\Omega_m^{(+1/2)}$ has been picked, \ie, 
after 
imposing equation \ref{eq:zcg}, this one
set must be used throughout. In particular, the right-hand sides of
 equations \ref{eq:zr} will also have to
be expressed in this basis. As the $\Omega_m^{(+1/2)}$ disappear 
from the computations  already at
the level of equation \ref{eq:zs}, the particular choice made is
irrelevant.  For the same reason, one can easily convince oneself
that the final PF equations obtained are independent of the actual
choice made.

Let us  point out two further consequences of this prescription used 
to define  $\tilde
M$. First, some of the equations collected in  \ref{eq:zs} may 
possibly involve,  both in the
right and the left-hand sides, occurrences of 
$\Omega_{n-1}^{(-1/2)}$
 and its modular derivatives,
as dictated by the prescription. One might worry that the latter will
 not disappear from the final
result for the SW differential $\lambda_{SW}$. That it  will always 
drop out follows from the fact
that none of the first $g$ equations in \ref{eq:zs} involves 
$\Omega_{n-1}^
{(-1/2)}$, as they are untouched
by the defining prescription of $\tilde M$. The decoupling procedure 
followed to
decouple $\lambda_{SW}$ also respects this property, as it basically
 discards all equations for the periods of 
non-holomorphic differentials (with the exception of the SW 
differential  itself, of
course). 

A second consequence of the prescription used to define  $\tilde M$
 is the fact that its
entries may now become {\it rational} functions of the moduli, 
rather  than {\it polynomial}
functions. This is  different from the situation for the $SO(2r+1)$,
$Sp(2r)$ and $SO(2r)$ gauge groups, where these entries were always
polynomials in the $u_i$. The reason is that  solving the linear
relation \ref{eq:zcg} for one particular $\Omega_m^{(+1/2)}$ may
 involve division by a polynomial in the $u_i$.

Having taken care of the difficulty just mentioned, \ie, the 
identification of the appropriate
space of periods on which  $\tilde M$ will be invertible, the rest 
of 
the decoupling
procedure already explained for the $SO(2r+1)$, $Sp(2r)$ and 
$SO(2r)$
 gauge groups holds throughout.
In particular, expressions totally analogous to those from equation
 \ref{eq:zaa} to \ref{eq:zag} continue to be
valid, with $s=g$. As a compensation for this technical difficulty 
of having nonzero odd Casimir
operators, one has that the SW differential truly becomes a 
potential for {\it all} holomorphic
differentials on the curve, so the equivalent of equation 
\ref{eq:zab}
 now  includes the odd holomorphic
periods as well.

\subsection{The exceptional gauge groups}

The method developed in  section 2 can also be applied to obtain 
the PF equations associated with
$N=2$ SYM  theories (with or without massless matter) when the 
gauge
group is an exceptional group, as the vacuum structure of these
theories is also described by hyperelliptic
curves \cite{ARDALAN,IRAN,DANII}. In principle, a set of 
(1st-order) PF
equations similar to
those given in equation
\ref{eq:zz} can also be derived. However, we have seen that the 
decoupling
procedure described in section 3 depends crucially on our ability to
identify an appropriate subspace of periods to restrict to. Such an
identification makes use of the  structure of the corresponding Weyl
group. With the exception of $G_2$, whose Weyl group is
$D_6$ (the dihedral group of order 12), the  Weyl groups of $F_4$,
 $E_6$, $E_7$ and $E_8$ are not
easily manageable, given their high orders. Therefore, we cannot
 hope to 
be able to develop a
systematic decoupling prescription similar to the one given for 
the classical gauge groups. This
is just a reflection of the exceptionality of the groups involved.

Another difficulty,  which we have illustrated in
the particular case of $G_2$ below, is the fact that the genus $g$ 
of the corresponding Riemann
surface $\Sigma_g$ will in general be too high compared with the 
number of independent Casimir
operators. As there is one modulus $u_i$ per Casimir operator, we 
cannot expect the SW differential
to be a  potential for {\it all} $g$  holomorphic differentials 
on $\Sigma_g$. This fact has
already been observed for the $B_r$, $C_r$ and $D_r$ gauge 
groups. However, the
 novelty here is that, in 
general, the best one can hope for
is to equate 
$\p\lambda_{SW}/\p u_i$ to some linear combination (with 
$u_i$-dependent coefficients) of a number of
holomorphic differentials $\omega_j$,
\renewcommand{\theequation}{3.\arabic{equation}}
\setcounter{equation}{26}

\be
{\p\lambda_{SW}\o\p u_i}=\sum_{j=0}^{g-1} c^j_i(u_l) 
\omega_j
\label{eq:zda}
\ee
Although the requirement of homogeneity with  respect to the
 (residual) $R$-symmetry can give us a
clue as to the possible terms that can enter the right-hand
 side of equation \ref{eq:zda}, the actual linear
combinations can only be obtained by explicit computation. In
 general, such linear combinations
may involve more than one non-zero coefficient $c^j_i(u_l)$. As a
 consequence, the decoupling
procedure 
 explained in previous sections breaks down, since it
 hinged on the SW differential
$\lambda_{SW}$ being a  potential for (some well 
defined subspace of)  the holomorphic 
differentials, \ie, on all $c^j_i(u_l)$
but one being zero. In other words, even if it were possible to
 identify the appropriate subspace of
periods that one must restrict to, the high value of the genus 
$g$ would probably prevent a
decoupling of the PF equations. 

As an illustration, we have included the details pertaining to 
$G_2$ in section 4.3.

\section{Examples}

\subsection{Pure $SO(5)$ SYM theory}

The vacuum structure of the effective,  pure $N=2$ SYM theory 
with 
gauge group $SO(5)$ is described
by the curve \cite{DANI}
\renewcommand{\theequation}{4.\arabic{equation}}
\setcounter{equation}{0}
\be
W=y^2= p(x)^2-x^2=x^8+2ux^6+(u^2+2t)x^4+2tux^2+t^2-x^2
\label{eq:zah}
\ee
where $p(x)=x^4+ux^2+t$. The quantum scale has been set to unity, 
 $\Lambda=1$, and the moduli
$u$ and $t$ can be identified with the second- and fourth-order 
Casimir operators of
$SO(5)$, respectively. The discriminant $\Delta(u, t)$ is given by
\be
\Delta(u, t)=256t^2(-27+256t^3+144tu-128t^2u^2-4u^3+16tu^4)^2
\label{eqn:zahi}
\ee
Equation \ref{eq:zah} describes a hyperelliptic Riemann surface of 
genus
$g=3$, $\Sigma_3$. The  holomorphic differentials on $\Sigma_3$ 
are $\d x/y$, $x\d x/y$ and $x^2\d
x/y$, while $x^4\d x/y$, $x^5\d x/y$ and $x^6\d x/y$ are  
meromorphic
 differentials of the second
kind. From 
equation \ref{eq:ze}, the SW differential is given by 
\be
\lambda_{SW}= (-3x^4-ux^2+t){\d x\o y}
\label{eq:zajii}
\ee
Both $p(x)$ and $\lambda_{SW}$ are even under
 $x \rightarrow -x$.\footnote{Recall that our convention leaves out
the $\d x$ term in  the differential.} We therefore restrict
 ourselves to the subspace of
differentials on $\Sigma_3$ spanned by 
$\{\d x/y, x^2\d x/y, x^4\d x/y, x^6\d x/y\}$. This is
further confirmed by the fact that the recursion relations 
\ref{eq:zn} and \ref{eq:zo} now have a step of 2 units,
\be
\Omega_n^{(-1/2)}={8\o n-3}\Big[t^2\Omega_n^{(+1/2)}+{3\o 4}
(2tu-1)\Omega_{n+2}^{(+1/2)}
+{1\o 2}(u^2+2t)\Omega_{n+4}^{(+1/2)}+{u\o 2}
\Omega_{n+6}^{(+1/2)}
\Big]
\label{eq:zaji}
\ee
and
\bea
\Omega_n^{(+1/2)}
& = & {1\o n-11}\Big[(10-n)2u\Omega_{n-2}^{(+1/2)}
+(9-n)(u^2+2t)
\Omega_{n-4}^{(+1/2)}\nonumber \\
&+&(8-n)(2tu-1)\Omega_{n-6}^{(+1/2)}+(7-n)t^2
\Omega_{n-8}^{(+1/2)}
\Big]
\label{eq:zaj}\eea
so that even and odd values don't mix. The solution of these 
recursions can be given in terms of the
initial data $\{\Omega_0^{(+1/2)}, \Omega_2^{(+1/2)}, 
\Omega_4^{(+1/2)}, \Omega_6^{(+1/2)}\}$,
where the indices take on the values allowed by the even subspace
 of differentials. From
equations \ref{eq:zaji} and \ref{eq:zaj}, the $M$ matrix of 
equation \ref{eq:zp} can be
 readily computed. Its determinant
is  found to be a product of powers of the factors of the
discriminant 
$\Delta(u,t)$:
\be
{\rm det}\, M={16\o 9} t^2(-27+256t^3+
144tu-128t^2u^2-4u^3
+16tu^4)
\label{eq:zal}
\ee
Therefore, it has the same zeroes as $\Delta(u,t)$ itself, but with 
different multiplicities.

Next, the change of basis in the space of differentials required by 
equation \ref{eq:zv} is effected by the
matrix 
\be
K=\pmatrix{1&0&0&0\cr
           0&1&0&0\cr
           t&-u&-3&0\cr
           0&0&0&1\cr}
\label{eq:zam}
\ee
The matrices $D(u)$ and $D(t)$ defined in equations \ref{eq:zq} 
and \ref{eq:zr} 
can  be computed  using the
expressions for $W$ and the 
recursion relations given in equations
  \ref{eq:zaji} and \ref{eq:zaj}.\footnote{For the sake of brevity, 
the explicit
expressions of these matrices are not reproduced here.}  Once
$D(u)$ and $D(t)$ are reexpressed  in the new basis $\{\pi_0, 
\pi_2, \pi_4=\lambda_{SW}, \pi_6\}$
defined by $K$ in \ref{eq:zam},  they produce the $U_i$
 matrices of 
equation \ref{eq:zy}. Let us just observe that,
from  the corresponding third rows of $U_t$ and $U_u$, 
one finds
\be
{\p\Pi_4\o\p t}=\Pi_0, \qquad 
{\p\Pi_4\o\p u}=\Pi_2
\label{eq:zan}
\ee
as expected for the SW period $\Pi_4=\Pi_{SW}$. 

Let us consider the $t$-modulus and carry out the decoupling
 procedure for the SW period
explicitly. Block-divide the $U_t$ matrix  as required by equation 
\ref{eq:zacc}:
\be
U_t=\pmatrix{A_t&B_t\cr C_t&D_t}
\label{eq:zao}
\ee
Specifically,  one finds
\bea
\lefteqn{A_t  =  {1\o {\rm det}\, M} {16t\o 27}}
\nonumber \\
& \left(
\begin{array}{cc}
t(-384t^2+27u-80tu^2+4u^4) &
(135t-240t^2u-27u^2+68tu^
3-4u^5)\\
4t^2(-18+76tu+u^3)&4t(60t^2-7tu^2-u^4)
\end{array}
\right)
\label{eq:zap}
\eea
\be
B_t={1\o {\rm det}\, M}
\left(
\begin{array}{cc}
-{64\o 27}t(12t^2+27u-47tu^2+4u^4)&
{16\o 3}t(27-48tu+4u^3)\\
-{16\o 27}t^2(9+160tu+16u^3)&{64\o 3}t^2(12t+u^2) 
\end{array}
\right) 
\label{eq:zaq}
\ee
We observe that 
\be
{\rm det}\, B_t = {256\o 9}t^3(27-256t^3-144tu+
128t^2u^2+4u^3-16tu^4)
\label{eq:zar}
\ee
so $B_t$ is invertible except at the singularities of moduli space.
 Carry out the
 matrix
multiplications of equation \ref{eq:zaf} to get
\be
\Pi_4=-(16t^2+{4\o 3}tu^2){\p\Pi_0\o\p t}+(9-16tu+
{4\o 3}u^3)
{\p\Pi_2\o\p t}-8t\Pi_0
\label{eq:zas}
\ee
Finally,  use \ref{eq:zan} to obtain a decoupled equation 
for the SW
 period $\Pi_4=\Pi_{SW}$:
\be
{\cal L}_1 \Pi_{SW}=0,\qquad 
{\cal L}_1=4t(u^2+12t){\p^2\o\p t^2}-(27-48tu+4u^3)
{\p^2\o\p t\p u}
+24t{\p\o\p t}+3
\label{eq:zat}
\ee
Analogous steps for the $u$ modulus lead to
\be
{\cal L}_2 \Pi_{SW}=0,\qquad 
{\cal L}_2=(
9-32tu){\p^2\o\p t\p u}-4(12t+u^2)
{\p^2\o\p u^2}-
8t{\p\o\p t}-1
\label{eq:zau}
\ee

\subsection{Pure $SU(3)$ SYM theory.}

The vacuum structure of the effective,  pure $N=2$ SYM 
theory with 
gauge group $SU(3)$ is described
by the curve \cite{LERCHE,ARG,KLEMM}\footnote{Our moduli 
$(u, t)$ correspond
to 
$(-u, -v)$ in \cite{LERCHE}.}
\be
W=y^2=p(x)^2-1=x^6+2ux^4+2tx^3+u^2x^2+2tux+t^2-1
\label{eq:zba}
\ee
where $p(x)=x^3+ux+t$. The quantum scale has been set to unity, 
 $\Lambda=
1$, and the moduli $u$
and $t$ can be identified with the second- and third-order Casimir 
operators of
$SU(3)$, respectively. The discriminant is given by
\be
\Delta(u, t)=-64(27-54t+27t^2+4u^3)(27+54t+27t^2+4u^3)
\label{eq:zbb}
\ee
Equation \ref{eq:zba} describes a Riemann surface of genus
$g=2$, $\Sigma_2$. The  holomorphic differentials  on  
$\Sigma_2$
are $\d x/y$ and $x\d x/y$, while
$x^3\d x/y$ and  $x^4\d x/y$ are meromorphic differentials of the
 second kind. From equation 
\ref{eq:ze}, the SW differential is given by 
\be
\lambda_{SW}= -(ux+3x^3){\d x\o y}
\label{eq:zbc}
\ee
The recursion relations \ref{eq:zn} and \ref{eq:zo} can be 
easily computed. They have a step of 1 unit,
\bea
\lefteqn{\Omega_n^{(-1/2)}  = } \nonumber \\
& {1\o n-2}\Big[4u\Omega_{n+4}^{(+1/2)}+6t
\Omega_{n+3}^{(+1/2)}+
4u^2\Omega_{n+2}^{(+1/2)}+10tu
\Omega_{n+1}^{(+1/2)}+6(t^2-1)
\Omega_n^{(+1/2)}\Big]
\label{eq:zbd}
\eea
and
\bea
\lefteqn {\Omega_{n}^{(+1/2)}= {1\o n-8}\Big[2u(7-n)
\Omega_{n-
2}^{(+1/2)
}+2t({13\o 2}-n)\Omega_{n-3}^{(+1/2)}}\nonumber \\
& +(6-n)u^2\Omega_{n-4}^{(+1/2)}+({11\o
2}-n)2tu\Omega_{n-5}^{(+1/2)}+(n-5)(1-t^2)\Omega_{n-6}
^{(+1/2)}\Big]
\label{eq:zbe}
\eea
which means that the solution for $\{\Omega_0^{(-1/2)},
\Omega_1^{(-1/2)}, \Omega_3^{(-1/2)},
\Omega_4^{(-1/2)}\}$ in terms of the initial values 
$\{\Omega_0^{(+1/2)}, \Omega_1^{(+1/2)},
\Omega_3^{(+1/2)}, \Omega_4^{(+1/2)}\}$ will involve both 
even and 
odd values of $n$. 
Inspection of the recursion relations above shows that, in the
 computation of $\Omega_4^{(-1/2)}$,
the value of $\Omega_8^{(+1/2)}$ is needed. This illustrates the 
situation described in section 3.3.

As already explained,  in order to properly identify the right
 subspace of periods to work with,
it suffices to include $\Omega_2^{(-1/2)}$ and its counterpart
 $\Omega_2^{(+1/2)}$.  It is then 
possible to use equations \ref{eq:zbd} and \ref{eq:zbe} in 
order to write a set of 4  equations  expressing
$\Omega_n^{(-1/2)}$, where $n=0,1,3,4$, in terms of 
$\Omega_n^{(+1/2)}$ with $n=0, 1, 2, 3,4$ {\it
and} $\Omega_2^{(-1/2)}$. As  prescribed in section 3.3, we pull 
all occurrences of
$\Omega_2^{(-1/2)}$ to the left-hand side. In the case at hand, 
such occurrences take place under
the form
\be
\Omega_4^{(-1/2)}=-u\Omega_2^{(-1/2)}+ {\rm terms}\; 
{\rm in}\;\Omega_m^{(+1/2)}
\label{eq:zbei}
\ee
while the expansions for the $\Omega_m^{(-1/2)}$, $m\neq 4$, 
do  not involve $\Omega_2^{(-1/2)}$.
We therefore define  $
\tilde M$  as the matrix of coefficients in
 the expansion of 
\be
\Omega_0^{(-1/2)},\quad
\Omega_1^{(-1/2)},\quad \Omega_3^{(-1/2)}, \quad
\Omega_4^{(-1/2)}+
u\Omega_2^{(-1/2)}
\label{eq:zdaa}
\ee
in terms of $\Omega_n^{(+1/2)}$ with $n=0, 1, 2, 3, 4$. 
Next we use
 the linear dependence relation
between the latter periods, as derived in equation \ref{eq:zcg}, 
which for
 the particular case at hand reads
\be
\Omega_2^{(+1/2)}=-{u\o 3} \Omega_0^{(+1/2)}
\label{eq:zdb}
\ee
We use  equation \ref{eq:zdb}  to express $\Omega_2^{(+1/2)}$ 
in terms of 
 $\Omega_0^{(+1/2)}$. This
choice has the computational advantage that, as the coefficicent of
 $\Omega_2^{(+1/2)}$ in
equation \ref{eq:zdb} is a constant, no division by a polynomial is 
involved, and the entries of $\tilde M$
continue to be polynomial functions in $u$ and $t$. From the above
  one can  immediately conclude that the periods of equation  
\ref{eq:zdaa} can be
expressed as certain linear combinations, with $u$ and $t$-dependent
coefficients, of
$\{\Omega_0^{(+1/2)}, \Omega_1^{(+1/2)}, \Omega_3^{(+1/2)}, 
\Omega_4^{(+1/2)}\}$. 
After this change of basis in the space of periods (or
 differentials) has been performed, the rest
of the construction already explained  goes through. We first
 confirm that $\tilde M$ so
defined  is invertible except at the singularities of moduli space,
 as
\be
{\rm det}\, \tilde M={4\o 9}(27-54t+27t^2+4u^3)
(27+54t+27t^2+4u^3)
\label{eq:zbf}
\ee
The SW differential is included  in 
the computations upon 
performing
 the change of basis from
$\{\omega_0,\omega_1, \omega_3, \omega_4+u\omega_2\}$
 to $\{\pi_1, 
\pi_2, \pi_3=\lambda_{SW}, \pi_4\}$, as given
by the matrix 
\be
K=\pmatrix{1&0&0&0\cr
           0&1&0&0\cr
           0&-u&-3&0\cr
           0&0&0&1\cr}
\label{eq:zbg}
\ee
where basis indices have also been relabelled for simplicity.
The matrices $D(u)$ and $D(t)$ defined in equations \ref{eq:zq} 
and \ref{eq:zr}  can be equally computed using the
expressions f
or $W$ and the recursion relations given above, and 
recast in the new basis
$\{\pi_1, \pi_2, \pi_3=\lambda_{SW}, \pi_4\}$ defined by $K$. 
This produces the $U_i$ matrices of
equation \ref{eq:zy}. Let us just observe that, from  the  third rows of
 $U_u$ and $U_t$, one finds
\be 
{\p\Pi_3\o\p t}=\Pi_1, \qquad {\p\Pi_3\o\p u}=\Pi_2
\label{eq:zbh}
\ee
as expected for the SW period $\Pi_3=\Pi_{SW}$. Further
carrying out the decoupling procedure as already prescribed yields
\be
{\cal L}_i\Pi_{SW}=
0,\qquad i=1,2
\label{eq:zbii}
\ee
where

\bea
{\cal L}_1 & = &(27+4u^3-27t^2){\p^2\o\p u^2}+12u^2t
{\p^2\o\p u\p t}+
3tu{\p\o\p t}+u \nonumber \\
{\cal L}_2 & = &(27+4u^3-27t^2){\p^2\o\p t^2}-36ut
{\p^2\o\p u\p t}-
9t{\p\o\p t}-3
\label{eq:zbj}
\eea
in complete agreement with \cite{KLEMM}, once the differences in 
notation have been taken into account.

\subsection{Pure $G_2$ SYM theory.}

The vacuum structure of the effective,  pure $N=2$ SYM
 theory with gauge group $G
_2$ is described
by the curve \cite{ARDALAN,IRAN,DANII}
\be
W=y^2=p(x)^2-x^4=x^{12}+2ux^{10}+{3u^2\o 2} x^8+
(2t+{u^3\o 2})x^6+
({u^4\o 16}+2tu-1)x^4+{tu^2\o
2}x^2+t^2
\label{eq:zea}
\ee
where $p(x)=x^6+ux^4+u^2 x^2/4+t$.\footnote{Some doubts 
about this
curve have recently been expressed in \cite{GIDDINGS}. The 
difficulty
encountered with equation (4.36) might perhaps be circumvented 
with a
different curve.} The quantum scale has been set to
 unity, \ie, $\Lambda=1$, and
the moduli $u$ a
nd $t$ can be identified with the second- and
 sixth-order Casimir operators of
$G_2$, respectively. We observe the absence of a fourth-order 
Casimir
operator; instead, its role is fulfilled by the {\it square} of the
second-order one. On the curve, this is reflected in the coefficient
of $x^2$. This is a consequence of the exceptionality of the Lie
algebra $G_2$. The discriminant $\Delta (u, t)$ is given by
\bea
\Delta(u, t)& = & 65536t^6
(-16+108t^2+72tu+8u^2-2tu^3-u^4)^2\nonumber \\         
             && ~~~~~~~~~~(16+108t^2-72tu+8u^2-
2tu^3+u^4)^2
\label{eq:zeb}
\eea
Equation \ref{eq:zea} describes a hyperelliptic Riemann 
surface of genus
 $g=5$, $\Sigma_5$. The
holomorphic differentials on $\Sigma_5$ are $x^j\d x/y$, where
 $j\in \{0,1,2,3,4\}$, while 
$x^{6+j}\d x/y$, with $j$ in the same range, are meromorphic
differentials of the second kind. From equation
\ref{eq:ze}, the SW differential is given by
\be
\lambda_{SW}=(2t-2ux^4-4x^6){\d x\o y}
\label{eq:zec}
\ee
Both $p(
x)$ and $\lambda_{SW}$ are even under
 $x\rightarrow -x$. We 
therefore restrict ourselves to
the subspace of differentials on $\Sigma_5$ spanned by 
$\{\d x/y$,
 $x^2\d x/y$, $x^4\d x/y$, $x^6\d
x/y$, $x^8\d x/y$, $x^{10}\d x/y\}$. This is further 
confirmed by 
the fact that the recursion
relations
\ref{eq:zn} and \ref{eq:zo} now have a step of 2 units,
\bea
\Omega_n^{(-1/2)} & = & {1\o n-5}\Big[12t^2 
\Omega_n^{(+1/2)}+5tu^2 
\Omega_{n+2}^{(+1/2)}
+(16tu+{1\o 2}u^4-8) \Omega_{n+4}^{(+1/2)}
\nonumber \\
&+ & (12t +3u^3)\Omega_{n+6}^{(+1/2)}+
6u^2\Omega_{n+8}^{(+1/2)}+4u \Omega_{n+10}
^{(+1/2)}\Big]
\label{eq:zed}
\eea
and
\bea
\Omega_n^{(+1/2)} & = & {1\o n-17}\Big[(11-n)t^2
\Omega_{n-12}^{(+1/2)}
+{1\o 2}(12-n)
tu^2\Omega_{n-10}^{(+1/2)}\nonumber \\ 
& + & (13-n)({u^4\o 16}+2tu-1)\Omega_{n-8}^
{(+1/2)}+(14-n)
(2t+{u^3\o 2})\Omega_{n-6}^{(+1/2)}\nonumber \\
&+ & (15-n){3\o 2} u^2 \Omega_{n-4}^{(+1/2)}
+(16-n)2u
\Omega_{n-2}^{(+1/2)}\Big]
\label{eq:zee}
\eea
so that even and odd values don't mix. The solution of these
 recursions can be given in terms of the
initial data $\{\Omega_0^{(+1/2)}, \Omega_2^{(+1/2)}, 
\Omega_4^{(+1/2)}, \Omega_6^{(+1/2)},
\Omega_8^{(+1/2)}, \Omega_{10}^{(+1/2)}\}$, where 
the indices take 
on the values allowed by the even
subspace of differentials picked above. From equations  
\ref{eq:zed} and
\ref{eq:zee}, the 
$M$ matrix of equation \ref{eq:zp} can be
readily computed. Its determinant is  found to be a 
product of
powers  of the factors of the discriminant
$\Delta(u,t)$:
\be
{\rm det}\, M={256\o
225}t^4(-16+108t^2+72tu+8u^2-2tu^3-u^4)
(16+108t^2-72tu+8u^2-2tu^3+u^4)
\label{eq:zef}
\ee
Therefore, it has the same zeroes as $\Delta(u,t)$ itself, but
 with different multiplicities.

Next, the change of basis  required by equation \ref{eq:zv} 
is effected 
by the matrix 
\be
K=\pmatrix{1&0&0&0&0&0\cr
              0&1&0&0&0&0\cr
              0&0&1&0&0&0\cr
              2t&0&-2u&-4&0&0\cr
              0&0&0&0&1&0\cr
              0&0&0&0&0&1\cr}
\label{eq:zeg}
\ee
The matrices $D(u)$ and $D(t)$ defined in equations 
\ref{eq:zq} and \ref{eq:zr} can be equally computed using the
expressions for $W$ and the recursion relations given in 
\ref{eq:zea}, 
\ref{eq:zed} and \ref{eq:zee}.\footnote{The complete system of 
coupled, first-order
equations is not reproduced here for the sake of brevity.}
Once $D(u)$ and
$D(t)$ are reexpressed  in the new basis $\{\pi_0, \pi_2,\pi_4,
 \pi_6=\lambda_{SW}, \pi_8,
\pi_{10}\}$ defined by $K$ in equation \ref{eq:zeg},  they 
produce the
 $U_i$ matrices of equation \ref{eq:zy}. Let us
just observe that, from  the corresponding fourth rows of $U_t$ 
and 
$U_u$, one finds
\be
{\p \Pi_6\o \p t}=\Pi_0, \qquad {\p \Pi_6\o \p u}=\Pi_4+
{u\o 2}\Pi_2
\label{eq:zeh}
\ee
Equation \ref{eq:zeh} is the expression of  \ref{eq:zda} when  
the gauge group is
$G_2$.  We observe
the presence of a linear combination of even holomorphic  
periods in
 the right-hand side, rather
than a clear-cut correspondence  between holomorphic 
periods 
(or differentials) and moduli. The
presence of an additional term $u\Pi_2/2$ is a consequence 
of the
 exceptionality of $G_2$. To
understand this fact in more detail, we observe that  $G_2$ has 
rank 2, so the number of
independent moduli is therefore 2. Like any other algebra, it has a
 quadratic Casimir operator
(associated with the $u$ modulus). In contrast to $SO(5)$, which 
is also rank 2, $G_2$ possesses no
fourth-order Casimir operator; instead, the next Casimir is of 
order 6. It is associated with the
$t$ modulus. There is no fourth-order Casimir operator other than
 the trivial one, namely, the one
obtained upon squaring the quadratic one. The existence of 
third- and fifth-order Casimir
operators is forbidden by the ${\bf Z}_2$ symmetry of the Weyl
 group. This leaves us with just 2
independent moduli, $u$ and $t$, to enter the definition of the
 curve as coefficients of $p(x)$.
As the latter is even and of degree 6 
(as dictated by the order
 of the highest Casimir operator),
we are clearly missing one modulus (associated with a would-be 
quartic Casimir), whose role is
then fulfilled by the square of the quadratic one. The consequence
 is a smaller number of moduli
(2) than would be required (3) for  the SW differential 
$\lambda_{SW}$ to serve as a potential for
the even holomorphic differentials.  This causes the presence
 of the linear combination in the
right-hand side of equation \ref{eq:zeh}, 
and the decoupling 
procedure 
breaks down.

\section{Summary and Outlook} 

In this paper an alternative derivation of the Picard-Fuchs equations
  has been presented which
 is systematic and  well suited for symbolic computer computations.
 It holds for any classical
gauge group, and aims explicitly at effective $N=2$ supersymmetric
 gauge  theories  in 4
dimensions. However, the techniques presented here may well find 
applications beyond these
specific areas. Our method makes use of the underlying 
 group theory 
in order to obtain a decoupled
set of partial, second-order equations satisfied by the period
 integrals of the Seiberg-Witten
differential. This computational simplicity allows one to derive the
 PF equations for large values of
the rank of the gauge group with comparatively very little effort. 
The inclusion of
 massless matter
hypermultiplets is also straightforward.
One of the strengths of the presentation of this paper is
 that the techniques studied
here lend themselves to a
 wide variety of applications, and are not 
limited to the SW problem only.

More interesting than the derivation of the PF equations themselves
 is of course the extraction of 
physical information from their solutions.  This topic has already
 been studied in the literature in
a number of cases 
\cite{KLEMM,BILAL,FERRARI,ITO,SASAKURA,THEISEN},
and provides an interesting  application of  our techniques, which we
are currently addressing \cite{NOS}. Another important extension of 
our
work  is 
the consideration of massive
matter hypermultiplets; this poses some technical challenges which
 are also under investigation
\cite{NOS}. We hope to be able to report on these issues in the near
 future.

\noindent{\bf Acknowledgements}

It's a great pleasure to thank Manuel Isidro San Juan for
 technical advice on the use of
{\sl Mathematica}. Discussions with him made the preparation of 
this paper much more enjoyable.

\newpage

\begin{center}
{\bf Appendix A}
\end{center}

\setcounter{equation}{0}
Below is a proof of the statement  that all zeroes of 
${\rm det}\, M$ are also zeroes of the discriminant 
$\Delta(u_i)$,
 possibly with different
multiplicities. For the sake of simplicity, we give the details 
pertaining to the gauge groups
$SO(2r+1)$, $SO(2r)$ and $Sp(2r)$ (with or without massless 
matter
in the fundamental representation).
 The proof for the $SU(r+1)$ gauge
groups is slightly more involved, but should be  technically 
analogous
to  the one presented below.

\renewcommand{\theequation}{A.\arabic{equation}}
To this purpose we recall equations \ref{eq:zzi} and 
\ref{eq:zzii}:
\be
\Delta (u_i)=a(x) W(x) + b(x) {\p W(x)\o\p x}
\label{eq:ya}
\ee 
\be 
{\phi(x)\o W^{\mu/2}}={1\o \Delta(u_i)}{1\o 
W^{\mu/2-1}}
\Big(a\phi+{2\o\mu-2}{\d\o\d
x}(b\phi)\Big)
\label{eq:yb}
\ee
As explained in the body of the paper, equation \ref{eq:yb} is an
 equivalent expression for  the
inversion of $M$, once it has been integrated along some closed
 1-cycle $\gamma\in H_1(\Sigma_g)$.
Let the polynomials $a(x)$ and $b(x)$  in equation \ref{eq:ya} 
have the 
expansions 
\be
a(x)=\sum_{i=0}^s a_ix^i, \qquad b(x)=\sum_{j=0}^{s'} b_jx^j
\label{eq:yc}
\ee
The respective degrees $s$ and $s'$ of $a(x)$ and $b(x)$  are 
easily found to be related by
$s'=s+1$. This follows from the fact that the left-hand side of 
equation \ref{eq:ya} has degree zero in
$x$. Moreover, $W(x)$ has degree $2n=2g+2$ in $x$. Using 
equation
\ref{eq:ya}, and imposing the condition
that the number of unknown coefficients  $a_i$ and $b_j$ equal 
the number of equations they must
satisfy, one easily finds $s=2g$ and $s'=2g+1$. It should be borne 
in mind that the coefficients
$a_i$ and $b_j$ themselves will be polynomial functions in the 
moduli $u_i$. For the gauge groups
$SO(2r+1)$, $SO(2r)$ and $Sp(2r)$ (with or without massless 
matter),
 we observe that
$a(x)$ will be even as a polynomial in $x$, while  $b(x)$
 will be odd. 

Let us take the polynomial $\phi(x)$ in equation \ref{eq:yb}
 to be
 $\phi(x)=x^m$, and set $\mu=3$ in equation \ref{eq:yb}
 to obtain the
periods $\Omega^{(\pm1/2)}$ as used in the body of the
 text.
Now assume the polynomial
$a(x)x^m+2(b(x)x^m)'$ has the following expansion in powers of
 $x$,
\be 
a(x)x^m+2{\d\o\d x}\big(b(x)x^m\big)=\sum_{r=0}
^{m+2g} 
c_r(u_i) x^r
\label{eq:yd}
\ee
where the $c_r(u_i)$ are some $u_i$-dependent coefficients. 
For the $SO(2r)$ and $SO(2r+1)$ gauge groups, $m$ can
 be assumed to
be even, while it can be assumed odd for $Sp(2r)$. 
Integration of
\ref{eq:yb}  along a closed 1-cycle
$\gamma\in H_1(\Sigma_g)$ produces
\be
\Omega_m^{(+1/2)}=-{3\o \Delta(u_i)}\sum_{r=0}
^{m+2g} c_r(u_i)
\Omega^{(-1/2)}_r
\label{eq:ye}
\ee
 As explained, equation \ref{eq:ye} can be taken to 
define the inverse $M$
 matrix, $M^{-1}$. In fact, it
just misses the correct definition by a minor technical point. 
As we 
let the integer $m$ run over 
(the even or the odd values of) the basic range $R$, the
 subscripts on
the
 right-hand side of equation \ref{eq:ye} will
eventually take values outside $R$. We can correct this 
by making 
 use of the recursion relation
\ref{eq:zmk}, in order to bring $m$ back into $R$. One 
just has to apply 
equation \ref{eq:zmk} for $\mu=-3/2$ and
substitute the  value of $k$ appropriate to the gauge group and 
matter content under
consideration.\footnote{For $N_f$ massless multiplets, one 
has $k=2+2N_f$
 for $SO(2r+1)$, $k=4+2N_f$ for
$SO(2r)$, and  $k=2(N_f-1)$ for $Sp(2r)$.} The resulting linear 
combination of periods 
$
\Omega_q^{(-1/2)}$ in the right-hand side has lower values 
of the 
subindex $q$.  One easily checks
that, as the degree of  $a(x)x^m+2(b(x)x^m)'$ is $2g+m$, the 
values
attained by $q$ in the right-hand side of equation \ref{eq:zmk} 
never  become
negative when $m$ runs over (the even or the odd subspace of)
$R$. Eventually, a linear combination (with $u_i$-dependent 
coefficients) will be obtained such that
all lower indices $q$ of the periods $\Omega_q^{(-1/2)}$ will lie 
within (the even or the odd subspace of)
$R$. Now this correctly defines an inverse $M$ matrix.

We have thus expressed the $(j,l)$ element of $M^{-1}$ as
\be
\Big[M^{-1}\Big]_{jl}={1\o \Delta(u_i)} P_{jl}(u_i) 
\label{eq:yg}
\ee
where the $P_{jl}(u_i)$ are certain polynomial functions of the
 moduli $u_i$. On the other hand,
from the definition of matrix inversion,
\be
\Big[M^{-1}\Big]_{jl}={1\o {\rm det}\, M} C_{jl}(u_i)
\label{eq:yh}
\ee
where $C_{jl}$ is the matrix of cofactors of $M$. Obviously,
 not all the $C_{jl}(u_i)$ are
divisible by
${\rm det}\, M$, as otherwise $M$ would be invertible even 
when 
${\rm det}\, M=0$.  From the
equality 
\be
{1\o {\rm det}\, M} C_{jl}(u_i) = {1\o \Delta(u_i)}
 P_{jl}(u_i)  
\label{eq:yi}
\ee
it follows that the right-hand side of equation \ref{eq:yi} 
will have
 to blow up whenever  the left-hand
side does, \ie, {\it all zeroes of ${\rm det}\, M$ are also 
zeroes 
of $\Delta(u_i)$}, possibly with
different multiplicities. The converse need not hold, as in
principle
 nothing prevents  all the
$P_{jl}(u_i)$ from simultaneously having $\Delta(u_i)$ as a
common 
factor.

Let us finally make an observation on the above proof for the 
$SU(r+1)$ gauge groups. From section 3.3
in the paper, we know that there are several different, though
 equivalent, ways of defining the
$\tilde M$ matrix. The particular choice made in solving 
the linear 
dependence relation \ref{eq:zcg} may
imply division by a non-constant polynomial $f(u_i)$ in the 
moduli
 $u_
i$, if the period that is
being solved for in equation \ref{eq:zcg} is multiplied by a 
non-constant
 coefficient. This has the effect
of causing a power of $f(u_i)$ to appear in the determinant 
${\rm det}\, \tilde M$, besides the
required powers of the factors of the discriminant 
$\Delta(u_i)$.   Obviously, only the zeroes of the
latter are relevant, as they are the ones associated with the 
singularities of the curve. The zeros
of ${\rm det}\, \tilde M$ due to the presence of $f(u_i)$ 
are a 
consequence of the prescription used
to define $\tilde M$.

\vfill
\break

\bigskip

\begin{center}
{\bf Appendix B}
\end{center}

Below are listed the PF equations satisfied by the period integrals
 of the SW differential of a
number of effective $N=2$ SYM theories (with and without 
massless matter), as  classified by their gauge groups.
Rather than an exhaustive list, we give a sample of cases in 
increasing order of the rank $r$ of the
gauge group, with some examples including massless matter 
hypermultiplets in the fundamental
representation. The  hyperelliptic curves describing their 
corresponding vacua are also quoted for
notational completeness. For computational simplicity, we 
systematically set the quantum scale
$\Lambda$ of the theory to unity, \ie,  $\Lambda=1$ throughout. 
The notation is as in the body of
the paper, \ie, the PF equations read 
$$ 
{\cal L}_i \Pi_{SW}=0, \qquad i=1,2, \ldots, r
$$ 
Wherever applicable, our results are coincident with those 
in the literature \cite{KLEMM,ITO,SASAKURA,THEISEN}.

$\bullet$ $N_f=0$ $SU(3)$

$$ 
y^2=(x^3+ux+t)^2-1 
$$
\beaa
{\cal L}_1 & = & (27+4u^3-27t^2){\p^2\o\p u^2}+12u^2t
{\p^2\o
\p u\p t}+3tu{\p\o\p t}+u\\
{\cal L}_2 & = & (27+4u^3-27t^2){\p^2\o\p t^2}-36ut
{\p^2\o\p u\p t}-9t{\p\o
\p t}-3
\eeaa

$\bullet$ $N_f=1$ $SU(3)$

$$ 
y^2=(x^3+ux+t)^2-x
$$

\beaa
{\cal L}_1 & = &\Big(-3t+{16u^3\o 45 t}\Big){\p\o\p t}\\
 & + & \Big(-9t^2-{5u^2\o 9t} + {28u^3\o
15}\Big){\p^2\o\p t^2}+\Big({25\o 4}-12tu+ {16u
^4\o 45 t}
\Big){\p^2
\o\p u\p t}-1\\
{\cal L}_2 & = & {1\over {(336ut - 100)}}{\Biggl\{} \Big[{300t} - 
{432 t^2 u}\Big]{\p\o\p t}
+   \Big[{-625}+  {3300tu}
 -  {3456t^2u^2}\Big]{\p^2\o\p t\p u}\\
& + &  \Big[{6480t^3}+{400u^2}-
{1344tu^3}\big]{\p^2\o\p u^2}{\Biggr\}} - 1\\
\eeaa

\newpage

$\bullet$ $N_f=2$ $SU(3)$

$$ 
y^2=(x^3+ux+t)^2-x^2
$$
\beaa
{\cal L}_1 & = & \Big(-3t-{8u\o 9t}+{8u^3\o 9t}\Big)
{\p\o\p t}+
\Big(-9t^2-{8u\o 3}+{8
u^3\o
3}\Big)  {\p^2\o\p t^2}\\
 & + &\Big({8\o 9t}-12 tu -{16u^2\o 9t}+{8u^4
\o 9t}\Big){\p^2\o\p t\p u}-1\\
{\cal L}_2 & = & -\big({3t\o u}+9tu\Big){\p^2\o\p t\p u}+
\Big(4+{27t^2 \o
 2 u}-4u^2\Big){\p^2\o\p u^2}-1
\eeaa

$\bullet$ $N_f=0$ $SO(5)$

$$ 
y^2=(x^4+ux^2+t)^2-x^2
$$
\beaa
{\cal L}_1 & = & 4t(u^2+12t){\p^2\o\p t^2}-(27-48tu+4u^3)
{\p^2
\o\p t\p u}+24t{\p\o\p t}+3\\
{\cal L}_2 & = & (9-32tu){\p^2\o\p t\p u}-4(12t+u^2)
{\p^2\o\p u^2}-8t{\p
\o\p t}-1
\eeaa

$\bullet$ $N_f=1$
 $SO(5)$

$$ 
y^2=(x^4+ux^2+t)^2-x^4
$$
\beaa
{\cal L}_1 & = & -16tu {\p^2\o\p t \p u}+(4-16t-4u^2)
{\p^2\o\p 
u^2}-1\\
{\cal L}_2 & = & -16tu
{\p^2\o\p t \p u}+(4t-16t^2-4tu^2){\p^2\o\p t^2}+
(2-8t-2u^2){\p\o\p 
t}-1
\eeaa

$\bullet$ $N_f=2$ $SO(5)$

$$ 
y^2=(x^4+ux^2+t)^2-x^6
$$
\beaa
{\cal L}_1 & = & (3t-{32\o 3}tu) {\p^2\o\p u\p t}+
(u-4u^2-{16t
\o 3}){\p\o\p u^2}+{8t\o
3}{\p\o\p t}-1\\
{\cal L}_2 & = & (3tu -16t^2-12tu^2){\p^2\o\p t^2}+
(3t -16tu+u^2-4u^3){\p^2\o\p t\p
u}+(2u
-8t-8u^2){\p\o\p t}-1
\eeaa

\vfill
\break

$\bullet$ $N_f=1$ $Sp(6)$

$$ 
y^2=(x^7+ux^5+sx^3+tx)^2-1
$$
\beaa
{\cal L}_1 & = & -\Big[24t-{32\o 7}su+{40\o 49}
u^3\Big]{\p\o\p
t}-\Big[8s-{16\o 7}u^2\Big] {\p\o\p s}\\
& + & \Big[-{147s\o t}
-36t^2+{16\o 7}stu-{20\o 49}tu^3\Big]
{\p^2\o\p t^2}\\
& + & \Big[-48st+{48\o 7}s^2u-{245u\o t}+
{4\o 7}tu^2-{60\o
49}su^3\Big]{\p^2\o\p s\p t}\\
& + & \Big[-16s^2-{343\o t }-24tu+{92\o
7}su^2-{100\o 49}u^4\big]{\p^2\o\p u\p t}-1\\
{\cal L}_2 & = & \Big[{294s\o t^2}+48t\Big]
{\p\o\p t}+\Big[-8s+{16\o
7}u^2\Big]{\p\o\p s}\\
& + & \Big[{441s^2\o t^2}+60st-{245u\o t}+{4\o
7}tu^2\Big]{\p^2\o\p t\p s}\\
& + & \Big[-16s^2-{343\o t}+{735\o t^2}su
+156tu+{12\o 7}su^2\Big]{\p^2\o\p s^2}\\
& + & \Big[{1029s\o
t^2}+252t-16su+{20\o 7}u^3\Big]{\p^2 \o\p u\p s}
-1\\
{\cal L}_3 & = & \Big[{294s\o t^2}+48t\Big]{\p\o\p t}+
\Big[-248s-{1764\o
t^3}s^2+{980\o t^2}u\Big] {\p\o\p s}\\
& + & \Big[-196s^2-{1323s^3\o
t^3}-{343\o t}+{1470su\o t^2}+156tu\Big]
{\p^
2\o \p t\p u}\\
& + & \Big[{1029s\o t^2}+252t-316su-
{2205s^2u\o t^3}+{1225u^2\o
t^2}\Big]{\p^2\o\p s\p u}\\
& + & \Big[-420s-{3087\o t^3}s^2+{1715\o
t^2}u-4u^2\Big]{\p^2\o\p u^2}-1
\eeaa


$\bullet$ $N_f=0$ $SO(7)$

$$ 
y^2=(x^6+ux^4+sx^2+t)^2-x^2
$$
\beaa
{\cal L}_1 & = & (-180t-268 su+{75u\o t}){\p^2\o\p u\p s}
+
(-100s^2+{25s\o t}-132tu){\p^2\o\p
u\p t}\\
& + & (-420 s+{125\o t}-4u^2){\p^2\o\p u^2}-24 t{\p\o\p t}
+
(-176s+{50\o t}){\p\o\p s} -1\\
{\cal L}_2 & = 
& (25-48st+{16\o 5}s^2 u-{4\o 5} tu^2-
{12\o 25}su^3){\p^2
\o\p t\p s}+ (-36t^2-{16\o 5}stu
+{12\o 25}tu^3){\p^2\o\p t ^2}-1\\
& + & (-16s^2-24tu+{52\o 5}su^2-{36
\o 25}u^4){\p^2\o\p t \p u}-24 t
{\p\o \p t}+({8\o 5}u^2-8s){\p\o\p s}-1\\
{\cal L}_3 & = & (-16s^2-
132 tu+{4\o 5} su^2){\p^2\o\p
s^2}+(25-84st-{4\o 5}tu^2){\p^2\o\p s \p t}\\
& + & (-180 t-16su +
{12\o 5}u^3){\p^2\o\p s\p u}-24
t{\p\o\p t} + ({8\o 5}u^2-8s){\p\o\p s}-1
\eeaa

$\bullet$ $N_f=1$ $SO(7)$

$$ 
y^2=(x^6+ux^4+sx^
2+t)^2-x^4
$$
\beaa
{\cal L}_1 & = & -(72t+64su){\p^2\o\p  u\p s}+
(16-16s^2-60tu)
{\p^2\o\p u\p t}\\
& - & (96s+4u^2){\p^2\o\p u^2}-  6t{\p\o\p t}-32 s
{\p\o\p s} -1\\
{\cal L}_2 & = & -(48st+2tu^2){\p^2\o\p t\p s}+
(-36
t^2-8stu+tu^3){\p^2\o\p t^2}\\
& + & (16-16s^2-24tu +8su^2
-u^4){\p^2\o\p t\p u} +(-24t-4su+{u^3\o 2})
{\p\o\p t}+(u^2-8s){\p\o\p
s}-1\\
{\cal L}_3 & = & (16-16s^2-60tu){\p^2\o\p s^2}-
(48st+2tu^2){\p^2\o\p s\p t}\\
& + & (-72t-16su+2u^3){\p^2\o\p s
\p u}-6t {\p\o\p t}+(u
^2-8s){\p\o\p s}-1
\eeaa


$\bullet$ $N_f=2$ $SO(7)$

$$ 
y^2=(x^6+ux^4+sx^2+t)^2-x^6
$$
\beaa
{\cal L}_1 & = & {2\o 9}(27-108t-48su+4u^3)
{\p\o\p t}-8s{\p\o\p
 s}+(9t-36t^2-16stu+{4\o 3}tu^3){\p^2\o\p t^2}\\
& + & (3s -48st-{16\o
3} s^2 u-4tu^2+{4\o 9} su^3){\p^2\o\p t\p s}\\
& + & (-16s^2-3u-24tu+4su^2-{4\o 9}u^4)
{\p^2\o\p u\p t}-1\\
{\cal L}_2 & = & -8s {\p\o\p s}-(36st+4tu^2)
{\p^2\o\p t\p s}\\ & - & (16s^2+36tu+{4\o
3}su^2){\p^2\o
\p s^2}+(9-36t-16su+{4\o 3}u^3)
{\p^2\o\p u\p s}-1\\
{\cal L}_3 & = & -8s
{\p\o\p s}-(4s^2+36tu) {\p^2\o\p t\p u}\\
& + & (9-36t-28su){\p^2\o\p u
\p s}-(36s+4u^2){\p^2\o\p u^2}-1
\eeaa

\vfill
\break

$\bullet$ $N_f=0$ $SU(4)$

$$ 
y^2=(x^4+sx^2+ux+t)^2-1
$$
\beaa
{\cal L}_1 & = & (s^2-8t){\p\o\p t}-3u{\p\o\p u}
+
(16-16t^2+3su^2){\p^2\o\p t^2}\\
& + & (7s^2u-24tu){\p^2\o\p u \p t}+(2s^3-16st-9u^2)
{\p^2\o\p s\p t}
-1\\
{\cal L}_2 & = & (s^2-8t){\p\o\p
t}-3u{\p\o\p u}+\Big(-{32s\o u}+{32st^2\o u}+
s^2u-24tu\Big){\p^2\o
\p t\p u}\\& + & (2s^3-16st-9u^2){\p^2\o\p u^2}+\Big(-{64\o u}
+{64t^2\o u} -12su
\Big){\p^2\o\p s \p u}-1\\
{\cal L}_3 & = & \Big(16t +{32s\o u^2}-{32st^2\o u^2}
\Big){\p\o\p t}-3u {\p\o
\p u}+\Big(32st+{64s^2\o
u^2}-{64s^2t^2\o u^2}-9u^2\Big){\p^2\o\p t\p s}\\
& + & \Big(-{64\o u}
+{64t^2\o u}-12su\Big){\p^2\o\p
u\p s}+\Big(-4s^2+96t+{128s\o u^2}-{128st^2\o u^2}
\Big) {\p^2\o\p 
s^2}-1
\eeaa

\vfill
\break

$\bullet$ $N_f=1$ $SU(4)$

$$  
y^2=(x^4+sx^2+ux+t)^2-x
$$
\beaa
{\cal L}_1 & = &  {1\over{({s^2} + 28t)}}{\Biggl\{} \Big[ {\left({4\over 7}{s^4}   + 8{s^2}t 
- 224 {t^2} + 27s{u^2}  \right)}{\p\o\p t} 
 +   {\left(   {9s^2u}-{84tu} \right)}
{\p\o\p u}\Big] \\
& + &    \Big[  {-{4\over 7}s^4t}-
{32s^2t^2} - {448t^3}
-{{147\over 4}su}+{120stu^2}\Big]
{\p^2\o\p t^2}\\
& + &     \Big[{{49\over 4}s^2} + {343t} + {{4\over 7}s^4u}
 +  {184s^2tu} - {672t^2u}
+{27su^3}\Big]{\p^2\o\p t\p u}\\
& + &   \Big[ {{12\over 7}s^5}+
{32
s^3t} - {448st^2} + {27s^2u^2} - {252tu^2}
\Big]{\p^2\o\p 
s\p t}{\Biggr\}} -1\\
{\cal L}_2 & = & {1\over{({160ut - 49})}} {\Biggl\{}  \Big[{-28s^2}+
{392t}+{112s^2tu}
-{704t^2u}\Big]
{\p\o\p t}  +  \Big[{{64\over 7}s^3t}+
{256st^2}+
{147u} -{480tu^2}\Big] {\p\o\p u}\\
& + &  \Big[{-2401\o 4}+
{1024s^3t^2\o 7}+{4096st^3}-{28s^2u}  +  {3136tu}+{112s^2tu^2}
-{3264t^2u^2}\Big]{\p^2\o\p t \p u}\\
& + &  \Big[{-84s^3}+{784st}+{{2112\over 7}s^3tu} -
{1792st^2u}
 +  {441u^2}- {1440tu^3}\Big]{\p^2\o\p u^2}\\
& + &   \Big[{64s^4t\o 7}+
{512s^2t^2}+
{7168t^3}
 +  {588su}-{1920stu^2}\Big]
{\p^2\o\p s\p u}{\Biggr\}} - 1
\eeaa

\beaa
{\cal L}_3 & = &  {1\over {(256s{t^2} - 49u + 196{u^2}t)}}{\Biggl\{}
 \Big[{2401\o 4} -
{6144st^3}  -  {3185 tu} +
{3136t^2u^2}
\Big]{\p\o\p t}\\ 
& + &  
\Big[{-196st} -
{128st^2u} 
 +  {147u^2}-{588tu^3}\Big]
{\p\o\p u}\\
& + &  \Big[{{7203\over 4} s} - {16384s^2t^3} - 
{9212stu} +  {6272st^2u^2}+
{441u^3} - {1764tu^4}
\Big]{\p^2\o
\p t \p s}\\
& + &  
\Big[{-196s^2t}-
{5488t^2}-{2432s^2t^2u}
 +  {17920t^3u}+
{588su^2} - {2352stu^3}\Big]
{\p^2\o\p u\p s}\\
& + & 
\Big[{16807\o 4}-
{1024s^3t^2} - {28672st^3}
 +  {196s^2u}-{21952tu}-{784s^2tu^2} + {228
48t^2u^2}\Big]
{\p^2\o\p s^2}{\Biggr\}}  - 1
\eeaa

\vfill
\break

$\bullet$ $N_f=2$ $SU(4)$

$$ 
y^2=(x^4+sx^2+ux+t)^2-x^2
$$
\beaa
{\cal L}_1 & = &  {1\over{({s^2} + 12t)}} {\Biggl\{} \Big[{\left( {-8s^2t}-
{96t^2}+
{27su^2}\right)}{\p\o\p t}
+  {\left( {9s^2u}-{36tu}\right)}
{\p\o\p u}\Big] \\
& + &   \Big[{-{4\over 3}s^4t}-{32s^2t^2}-{192t^3} + 
{72stu^2}\Big]
{\p^2\o\p t^2} +    \Big[{-27su} + {72s^2tu}-{288t^2u}+
{27su^3}
\Big]{\p^2\o\p t\p u}\\
& + &   \Big[{9s^2}+{4s^5\o 3}+
{108t}-
{192st^2}+{27s^2u^2}-{108tu^2}\Big]
{\p^2\o\p t\p s}{\Biggr\}}  - 1\\
{\cal L}_2 & = & \Big[{s^2\o 2}-2t\Big]{\p\o\p
t}+\Big[{2s^3\o 9u}+{8st\o 3u}-3u\Big]{\p\o\p u}\\
& + & \Big[{-s^2\o 2u}-{6t\o u}+{16s^3t\o
9u}+{64st^2\o 3u}+{s^2u\o 2}-18tu\Big]
{\p^2\o \p t\p u} +
(9+2s^3-8st-9u^2){\p^2\o\p u^2}\\
& + & \Big[{2s^4\o 9u}+{16s^2t\o 3u}+{32t^2\o u}
-12 su\Big]
{\p^2\o\p s\p u
} - 1\\
{\cal L}_3 & = &  {1\over{(-9 + 32st + 9{u^2})}}{\Biggl\{} \Big[
{\left ({72t}-{256st^2}+{144tu^2}
\right)} {\p\o\p t} 
  +  {\left( {27u}-{27u^3}\right)}
{\p\o\p u}\Big]\\
& + &  \Big[{-81} + {576 st} - 
{1024s^2t^2}
 +  {162u^2}+{288stu^2}-{81u^4}\Big]
{\p^2\o\p t \p s}\\
& + &  \Big[{108su}-{288s^2tu}+
{1152t^2u}
-{108su^3}\Big]{\p^2\o\p u\p s}\\
& + &  \Big[{36s^2}
+{432t}-{128s^3t} -  {1536st^2}-{36s^2u^2}+{1296tu^2}\Big]{\p^2\o\p
s^2}{\Biggr\}} - 1
\eeaa

\end{document}